# Multiplicative Component GARCH Model of Intraday Volatility

Xiufeng Yan

Phd, University of Sussex, 2021

School of Mathematical and Physical Sciences

University of Sussex

BN1 9QH

**Abstract.**

This paper proposes a multiplicative component intraday volatility model. The intraday conditional volatility is expressed as the product of intraday periodic component, intraday stochastic volatility component and daily conditional volatility component. I extend the multiplicative component intraday volatility model of Engle (2012) and Andersen and Bollerslev (1998) by incorporating the durations between consecutive transactions. The model can be applied to both regularly and irregularly spaced returns. I also provide a nonparametric estimation technique of the intraday volatility periodicity. The empirical results suggest the model can successfully capture the interdependency of intraday returns.

**Table of content**

# List of Tables

## List of Figures

# Multiplicative Component GARCH Model of Intraday Volatility

1. **Introduction**

The modelling of return dynamics is of both theoretical and empirical importance for financial research. Among the vast literatures on this topic, the ARCH-GARCH models are widely developed and are accepted as the standard techniques for the analysis of return volatility (Engle 1982 and Bollerslev 1986). The ARCH-GARCH models are developed to describe the phenomena of volatility clustering that is the positive autocorrelation in return volatility. However, financial research using high frequency data generally suggest the inadequacy of this standard time series volatility model for the modelling of intraday return dynamics. In general, when applied to high frequency intraday returns, estimations of GARCH models are often contradictory and defy theoretical predictions. For instance, Andersen and Bollerslev (1997) identified that the estimation using GARCH model for returns with different intraday frequencies gave parameters that are inconsistent with the theoretical results on time aggregation of GARCH estimation from Dorst and Nijman (1993).

It is well known that the return volatility systematically varies over the trading day and are highly correlated with the variation of other random variables associated with the transaction records such as the volumes and waiting-time over the trading day. In many markets, this systematical variation of intraday return volatilities is identified as a U-shape pattern in return volatility over the day. Specifically, returns during market opening and closing hours are much more volatile. Such intraday periodic pattern in the return volatility in equity market has been proven to have strong influence on the dynamics of equity returns (Bollerslev 1994, Guillaume et al. 1995). Consequently, in order to carry out meaningful analysis about intraday volatility dynamics, it is necessary to consider the intraday periodic volatility pattern as a fundamental determinant of the intraday volatility process.

In this paper, I investigate the difficulties encountered by GARCH volatility models when applied to the high frequency returns. I focus on the strong impact of intraday periodicity of return dynamics on the modelling of intraday volatility process using standard GARCH models. Besides, I construct our estimation of the intraday periodicity in volatility process based on the realized volatility calculated from tick-by-tick transactions. The estimation procedure could be applied to both time-aggregated and transaction-aggregated returns. The periodicity estimation based on tick-by-tick returns provide the most accurate description of intraday periodic volatility pattern. I also analyze the different impact of the intraday periodic feature of return dynamics on returns sampled from both fixed time interval and fixed number of transactions at various frequencies to verify our estimation of the intraday periodicity of return dynamics. Moreover, to identify the distinctive characteristics of the intraday returns process, the findings are compared with the corresponding features of daily returns series. Finally, I generalize the multiplicative GARCH model for intraday volatility of Engle (2012). I modify the intraday volatility periodicity component of the model to allow it to be conditional on the varying trading period of tick-by-tick transactions.

This paper is organized as follows. Section 2 presents the discussion about literatures regarding the intraday volatility modelling. Section 3 describes the data set and presents the intraday return pattern. Section 4 exhibits the multiplicative component intraday volatility models. Section 5 discusses the specification of the multiplicative volatility model using empirical evidences. It also includes my characterization of intraday volatility periodicity pattern. A relatively simple model that allows the estimation of intraday volatility periodicity to be conditional on the trading period. Section 6 contains the empirical analysis regarding our multiplicative component volatility models. I also analyze the statistical properties of the corresponding filtered returns series generated by normalizing the raw return series according to the estimated intraday volatility periodicity and the corresponding estimated daily volatility. I employ high frequency time-aggregated returns at different frequencies. Estimates of the degree of volatility persistence at the various sampling frequencies are compared with the theoretical aggregation results. Section 7 presents the conclusion.

## 2. Literature Review

Early in 1980s, authors like Wood et al. (1985) and Harris (1986) gave the empirical evidence about the distinctive U-shaped pattern in return volatility over the trading day using average intraday returns of stock market. After that, the intraday periodic feature of asset returns is widely reported and considered as a stylized fact regarding financial return series in different markets. For instance, similar evidence for foreign exchange markets can be found in Muller et.al (1990) and Baillie and Bollerslev (1991). Meanwhile, a main topic of financial research using high frequency data is intraday return volatility modelling based on the ARCH-GARCH model of Engle (1982) and Bollerslev (1986). These are partially motivated by an attempt to identify the economic origins of the volatility clustering phenomenon such as the mixture of distributions hypothesis; see for example Clark (1973), Tauchen and Pitts (1983), Harris (1987), Gallant et al. (1991), Ross (1989) and Andersen (1994, 1996).

A direct comparison of these research is difficult because of the different sampling frequencies used by these literatures. Nevertheless, as argued by Ghose and Kroner (1984), Guillaume (1994), Andersen and Bollerslev (1997) and Engle (2012), most of the results regarding degree of volatility persistence and parameter estimation conflicts with the theoretical aggregational results for ARCH-GARCH models by Nelson (1992) and Dorst and Nijman (1993). A plausible explanation is that the strong intraday volatility periodicity cannot be captured by GARCH-ARCH models. Specifically, the intraday volatility periodicity causes the contradictory phenomena between estimation of GARCH-ARCH parameters for returns at different intraday frequencies and theoretical predictions.

In 1990s, efforts on the modelling of intraday volatility periodicity pattern are given by the research group at Olsen and Associates on the foreign exchange market. For instance, Muller et al. (1990,1993) and Dacorogna et al. (1993) use time invariant polynomial functions to approximate the trading activities in different FOREX exchange market during the 24-hour trading cycle.

Andersen and Bollerslev (1997) gave a more general methodology for the modelling of intraday volatility periodicity pattern. They construct a multiplicative model of daily and intraday volatility for the 5-minute returns on both Deutschemark-dollar exchange rate and US stock market. They express the conditional variance as a product of intraday component and daily component. This specification allows the periodic pattern to be conditional on the current overall level of return volatility. Other closely related models for intraday volatility periodicity can be found in Ghose and Kroner (1996), Andersen and Bollerslev (1998), and Giot (2005). For instance, Giot (2005) adds a deterministic intraday periodicity pattern to the GARCH (1,1) and the EGARCH (Nelson 1991) models and estimates the two model for high frequency return series. Taylor and Xu (1997) provided an alternative specification for high frequency return volatility. They construct an hourly volatility model using an ARCH specification. The conditional variance specification is modified with two elements: the implied volatility and the realized volatility. In 2012, Engle developed a multiplicative component GARCH models based on the work of Andersen and Bollerslev (1998). Specifically, the conditional variance of 10-minute return series in US stock market is expressed as a product of daily, intraday periodicity and stochastic intraday volatility components. Compared with the model of Andersen and Bollerselv (1998) that considers the intraday volatility pattern as deterministic, Engle used two intraday components for the conditional variance model: a deterministic diurnal pattern and stochastic intraday ARCH.

Most of the discussed above use returns sampled from fixed time interval. Consequently, the estimation of intraday volatility periodicity is also based on observations that are equally spaced in time with the same length of interval. Meanwhile, a fundamental feature of tick-by-tick transaction records is that observations are not equally spaced in time. Returns sampled from fixed number transactions exhibit different statistical characteristics. For instance, the distribution of transaction-aggregated returns approaches the normal distribution much faster than the time-aggregated returns as the frequencies of returns decrease. Thus, it is interesting to examine whether the models based on time-aggregated returns can be extend to the transaction-aggregated returns, specifically, whether the intraday volatility periodicity estimation based on realized volatility from previous literatures such as Anderson and Bollerslev (1997) can be extended to the transaction-aggregated returns. Finally, I want to estimate the intraday periodicity directly from the tick-by-tick transaction records instead of imposing fixed observing time interval.

Based on the work of Anderson and Bollerslev (1997) and Engle (2012), I use the tick-by-tick transactions to estimate the intraday volatility periodicity component and allow the estimation to be conditional on the trading period during the day. Thus, it can be applied to transaction-aggregated return series and should capture the intraday volatility periodicity more accurately. My estimation method different from the previous research on intraday volatility modelling mainly in three perspectives. First, I generalize the multiplicative component intraday volatility model of Engle (2012) for tick-by-tick return sample. Second, my estimation of intraday periodicity is based on the tick-by-tick transactions, it should give more accurate description on the intraday volatility pattern in the specific time durations. Finally, I test the specification using both time-aggregated and transaction-aggregated returns.

3. **Data**

In this section, I give the data set. In order to comprehensively illustrate the periodic pattern of returns during the trading day, I focus on the time-aggregated return series. In Section 3.1, I first exhibit the descriptive statistics about the time-aggregated return series at various frequencies. I focus on the *5-minute* and *10-minute* return series. There are three main reasons for choosing the two frequencies. First, since most of the discussed literatures used these frequencies, I wish to compare the intraday periodic pattern of my data sample with theirs. Second, the preliminary tests show that the intraday pattern is not obvious for returns on higher frequency such as 30s, 1 min and 3 min. Third, the estimation of intraday periodic pattern during short time interval rises computation difficulties. In Section 3.2, I exhibit the intraday periodic pattern of the *5-minute* and *10-minute* return series.

## 3.1 Descriptive statistics

The data set is the tick-by-tick transaction records of symbol SPY during 01/02/2014-12/31/2014. It contains transactions for 252 trading days. The out-of-hour transactions whose occurrence lie outside the time period between 9:30 am and 4:00 pm on each day are removed.

|  | Sample Size | Mean | Variance | Kurtosis | Skewness | Maximum | Minimum |
|---|---|---|---|---|---|---|---|
| **30Sec** | 196560 | 1.29618E-07 | 5.30347E-08 | 264.8727029 | -0.173806698 | 0.01225414 | -0.012983101 |
| **1Min** | 98280 | 2.59236E-07 | 9.67974E-08 | 95.91646403 | -0.103161028 | 0.012458815 | -0.012906394 |
| **1.5Min** | 65520 | 3.88855E-07 | 1.39691E-07 | 80.86530506 | -0.005621653 | 0.012036628 | -0.013264313 |
| **3Min** | 32760 | 7.77709E-07 | 2.70802E-07 | 39.85096976 | -0.238157281 | 0.011012149 | -0.014459855 |
| **5Min** | 19656 | 1.29618E-06 | 4.06658E-07 | 8.436150587 | 0.075305595 | 0.006182118 | -0.005359357 |
| **10Min** | 9828 | 2.59236E-06 | 8.03275E-07 | 7.918815602 | -0.037720742 | 0.008269006 | -0.006320998 |
| **15Min** | 6552 | 3.88855E-06 | 1.17727E-06 | 7.388130566 | -0.079263515 | 0.007751481 | -0.006653523 |
| **30Min** | 3276 | 7.77709E-06 | 2.34033E-06 | 6.660810707 | -0.095190193 | 0.011082252 | -0.009269532 |
| **78Min** | 1260 | 2.02204E-05 | 6.20987E-06 | 6.687141886 | -0.195275515 | 0.014103076 | -0.014359004 |

Table 1. Descriptive statistics of time-aggregated returns

In this section I focus on the 5-minute and 10-minute return series. It seems natural to choose the time-aggregated return to present an illustration of intraday periodic pattern of the returns over the trading day. I choose the absolute value of the returns and squared returns as the proxies of the return volatility.

|  | Mean | Variance | Kurtosis | Skewness | Maximum | Minimum | First Lag autocorrelation |
|---|---|---|---|---|---|---|---|
| **5-Minute Return** | 1E-06 | 4.07E-07 | 8.4362 | 0.0753 | 0.006182 | -0.0054 | -0.015058796 |
| **Absolute 5-Minute Return** | 0.0004 | 2.13E-07 | 14.733 | 2.5577 | 0.006182 | 0 | 0.256396005 |
| **Squared 5-Minute Return** | 4E-07 | 1.23E-12 | 237.48 | 11.289 | 3.82E-05 | 0 | 0.179436501 |

Table 2. Descriptive statistics of 5-minute return series, absolute 5-minute return series and squared 5-minute return series

|  | Mean | Variance | Kurtosis | Skewness | Maximum | Minimum | First Lag autocorrelation |
|---|---|---|---|---|---|---|---|
| **10-Minute Return** | 2.59E-06 | 8.03E-07 | 7.918815602 | -0.03772074 | 0.008269006 | -0.006321 | -0.02467549 |
| **Absolute 10-Minute Return** | 0.000621954 | 4.16E-07 | 13.50275706 | 2.470604056 | 0.008269006 | 0 | 0.247384273 |
| **Squared 10-Minute Return** | 8.03E-07 | 4.46E-12 | 197.0880086 | 10.11289837 | 6.84E-05 | 0 | 0.16996921 |

Table 3. Descriptive statistics of 10-minute return series, absolute 10-minute return series and squared 10-minute return series

The mean of the 5-minute return series and 10-minute return series are 1E-06 and 2.59E-

06 respectively. Given their variance of 4.07E-07 and 8.03E-07, the mean of 5-minute return series and 10-minute return series can safely assumed to be zero. However, as suggested by the kurtosis in Tables 2 and 3, they are clearly not normally distributed. An interesting fact is that although the 5-minute return series have a positive skewness of 0.0753, the 10-minute return series have a negative skewness of -0.0378. Meanwhile, the maximum and minimum of the 5-minute returns are 0.0062 and -0.0054 respectively. These values do not suggest of sharp discontinuities in the return series. The maximum and minimum of 10-minute return series exhibit similar characteristics.

Note that the first lag autocorrelations of the 5-minute and 10-minute returns, -0.0151 and -0.0247 respectively, are significant since corresponding lower confidence bounds at the significance level of 5% are -0.0143 and -0.0201 respectively. The values of the correlation are nearly negligible from the economic perspective. In contrast, the volatilities of the 5-minute returns and 10-minute returns have highly significant positive autocorrelations. For instance, the absolute 5-minute return series has a first lag autocorrelation of 0.2564 and the absolute 10-minute return series has a first lag autocorrelation of 0.2474. The upper confidence bounds of the 5-minute return series and the 10-minute return series at the significance level of 5% are 0.0143 and 0.0201 respectively. The positive correlations of return volatility clearly suggest the phenomena of volatility clustering. The squared returns exhibit slightly smaller but also highly significant positive first lag autocorrelations.

### 3.2 Intraday return periodicity

In order to evaluate the intraday volatility periodicity, I present the plot of average and absolute average during the trading day for both 5-minute return series and the 10-minute return series first.

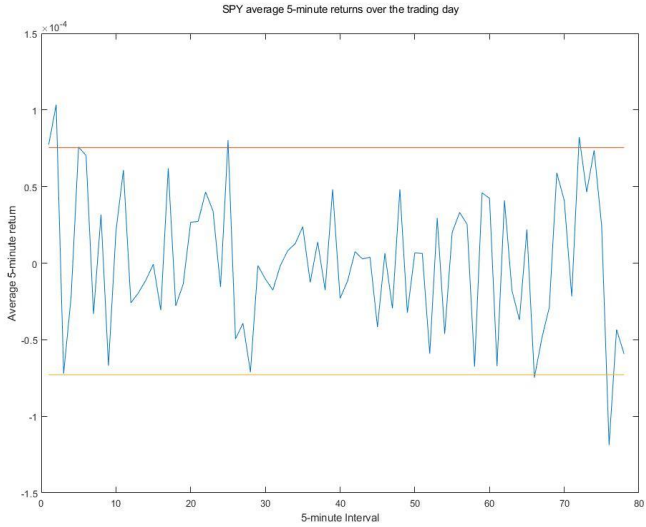

**Figure 1. The average SPY 5-minute return over the trading day**

The average SPY 5-minute returns are centered around 0 but fluctuate dramatically during the trading day. Figure 2 suggests that the average 5-minute returns during the opening and closing hours of the trading day are much more volatile. For instance, three of the four violations of the constant 5% confidence band for the null hypothesis of an *iid* series occur during the first and last ten intervals. Specifically, the first ten interval represents the period from 9:30 to 10:20 and last ten interval represents the period from 3:10 to 4:00. The sharp drop off during such intervals provide graphic evidence. Also note that, the Andersen-Darling statistics cannot reject the null hypothesis that the average 5-minute returns come from a normal distribution at significance level of 5%.

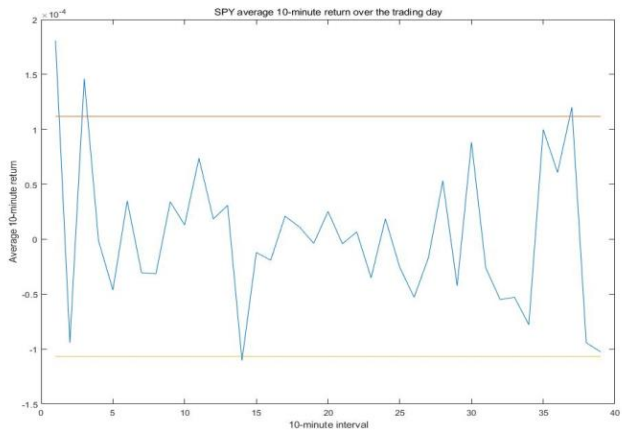

**Figure 0. The average SPY 10-minute return over a trading day**

Figure 2 gives similar but more clear presentation of intraday return periodicity for the average 10-minute return series. Three of the four violations of the constant 5% confidence band for the null hypothesis of an *iid* series occur during the first and last five intervals, which is consistent with the finding of the average 5-minute return series. As the case of average 5-minute returns, the average 10-minute returns also can be safely assumed to be normally distributed. It is worthy to mention that during the end of trading day, the dramatic variation of SPY 5-minute returns in Figure 1 no longer exist in Figure 2. In contrast, the 10-minute SPY average returns tend to be more volatile than 5-minute SPY average return during the opening hours.

Figure 3 exhibits the volatility (measured as the absolute average returns) of SPY average 5-minute returns during the trading day. It clearly verifies the pronounced U-shape of return volatility in financial markets. The volatility ranges from a low of around 0.03% at the 13:05 to a high of around 0.08% at the 9:35. It starts out at a relatively high level of 0.07% at 9:30 and decays slowly to 0.03% at 13:05, and then it again surges to around 0.05% at 16:00. The highly significant first lag positive autocorrelation of 0.256 in Table 2 also support the existence of intraday periodic volatility pattern. Figure 4 suggests similar periodic pattern for 10-minute return volatility. It also can be concluded from Figures 3 and 4 that the returns are more volatile during the opening hours than the closing hours, which corresponds to the 10am FED news announcements.

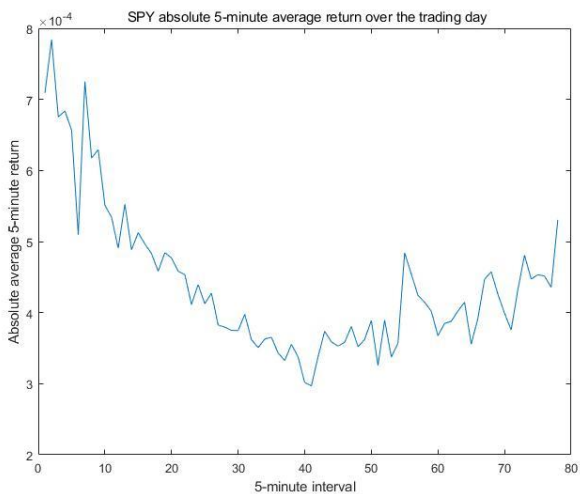

**Figure 1. The absolute average SPY 5-minute return over a trading day**

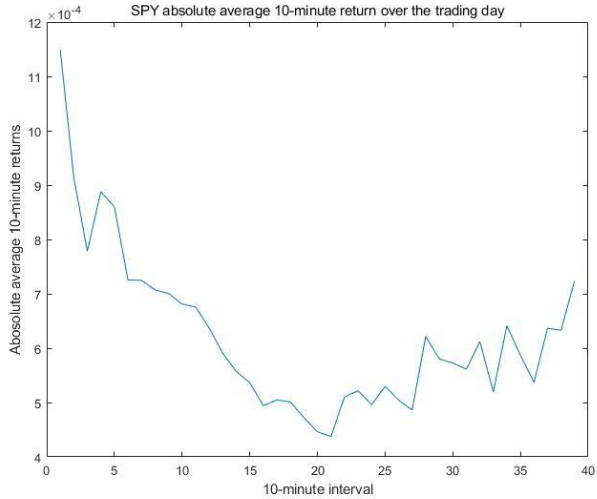

**Figure 4. The absolute average SPY 10-minute return over a trading day**

To support my prior investigation of the unconditional intraday volatility periodic pattern, I now examine the autocorrelations of the absolute 5-minute and absolute 10-minute return series to explore the dynamic feature of those two return series.

|       | 5-minute return | absolute 5-minute return | 10-minute return | absolute 10-minute return |
|-------|-----------------|--------------------------|------------------|---------------------------|
| Lag1  | -0.015058796    | 0.256396005              | -0.024675493     | 0.247384273               |
| Lag2  | -0.02727115     | 0.27393497               | -0.007165027     | 0.262341418               |
| Lag3  | 0.011513466     | 0.266731377              | 0.025501631      | 0.271487025               |
| Lag4  | -0.002240064    | 0.25614196               | -0.018678124     | 0.256228375               |
| Lag5  | -0.001991967    | 0.248556941              | 0.038864765      | 0.246937983               |
| Lag6  | 0.007446405     | 0.267334076              | 0.029325701      | 0.236207573               |
| Lag7  | 0.005399493     | 0.265061453              | 0.00940831       | 0.230465298               |
| Lag8  | -0.004431232    | 0.230266141              | -0.002262395     | 0.217919076               |
| Lag9  | 0.004174203     | 0.244515482              | -0.002524387     | 0.209070546               |
| Lag10 | 0.023174997     | 0.230559214              | 0.02308969       | 0.199440976               |
| Lag11 | 0.022193775     | 0.232530198              | 0.008302441      | 0.189182483               |
| Lag12 | 0.00899738      | 0.219332721              | -0.015269527     | 0.188770108               |
| Lag13 | -0.005247707    | 0.217748927              | -0.011524974     | 0.185738101               |
| Lag14 | 0.013262886     | 0.222211507              | 0.025102784      | 0.194993096               |
| Lag15 | 0.007466957     | 0.219513545              | 0.009805094      | 0.174629519               |
| Lag16 | -0.009635652    | 0.199277327              | 0.004593636      | 0.175036007               |
| Lag17 | -0.004045313    | 0.214672245              | -0.027126605     | 0.155205843               |
| Lag18 | 0.006474428     | 0.200850752              | -0.029243554     | 0.152888546               |
| Lag19 | 0.00035781      | 0.184185116              | -0.017268827     | 0.145260443               |
| Lag20 | 0.009019369     | 0.18152872               | 0.010949894      | 0.141655001               |
| Upper Bounds | 0.01426535 | 0.01426535 | 0.020174251 | 0.020174251 |
| Lower Bounds | -0.01426535 | -0.01426535 | -0.020174251 | -0.020174251 |

Table 4. Autocorrelations of average returns and absolute average returns

As discussed previously, the 5-minute return series and 10-minute return series can be plausibly considered as uncorrelated. For instance, although 5-minute return has significant first and second lag autocorrelations, the values of the autocorrelations are as small as -0.015 and -0.027 respectively. In contrast, the absolute return series exhibits highly significant strong positive correlations at almost all 20 lags. Observe that the autocorrelation of both absolute 5-minute and absolute 10-minute return series decay very slowly. If there is a periodic pattern of the volatility during each trading day, the correlations should also have a 24-hour cycle pattern.

To further investigate the intraday volatility periodicity pattern, I plot the correlogram for both absolute 5-minute return and absolute 10-minute return series for up to five days.

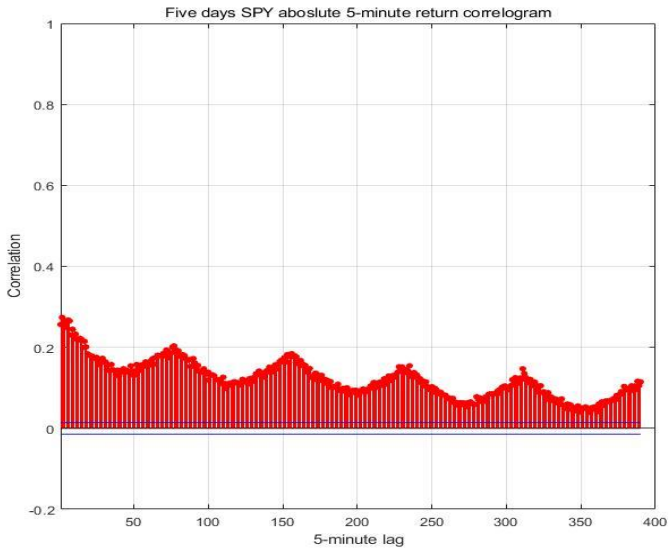

Figure 5. Five days SPY absolute 5-minute return correlogram

The Figures 5 and 6 present autocorrelation pattern of the absolute returns. The intraday volatility periodicity is exhibited as the corresponding U-shape in the correlogram. The correlations at daily frequency decay very slowly during the five days. Observe that the slowly declining U-shape cycles around every 80 intervals for the 5-minute lag and every 40 intervals for the 10-minute lag. Since the NYSE operates from 9:30 to 4:00 on each trading day, U-shape cycles on daily frequency exactly.

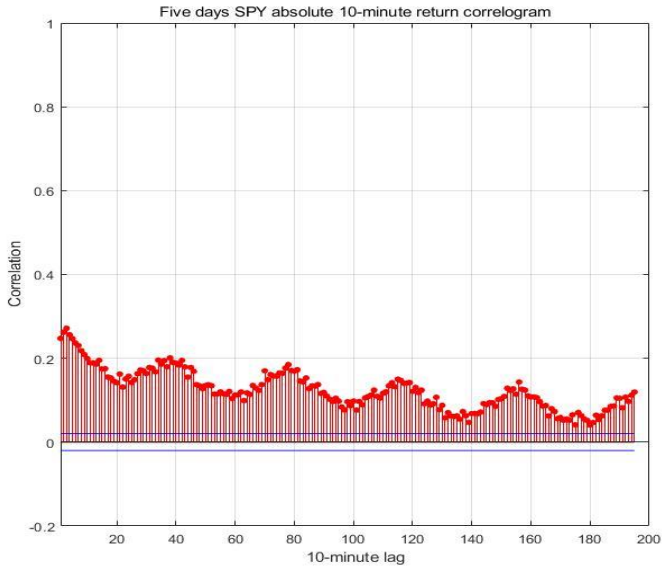

**Figure 6. Five days SPY absolute 10-minute return correlogram**

## 4 Multiplicative component intraday volatility model

In this section I present my multiplicative component intraday volatility model. The model is based on the work of Andersen and Bollerslev (1997) and Engle (2012). In order to illustrate the reasoning behind the model specification, I first discuss the interpretation of the intraday volatility periodicity pattern in section 4.1. I then present the model and discuss the econometric issues in section 4.2.

### 4.1 Interpretation of the intraday return volatility

The periodic pattern of the intraday volatility discussed in Section 3 suggests that a direct use of the ARCH-GARCH for modelling of the intraday return volatility could be inappropriate. The main reason is that the standard ARCH-GARCH models impose a geometric decay for the return autocorrelation structure. It cannot capture the periodic pattern of the volatility correlation that are reported in Section 3. In order to jointly describe the intraday periodicity and daily conditional heteroskedasticity, Andersen and Bollerslev (1997) present the following stylized specification of the intraday returns. Consider an intraday time-aggregated logarithm return series $\{r_{t,n}\}_{t,n}$, where $r_{t,n}$ represent the $n$th return at day $t$. Suppose there are $N$ observation during each trading day. Then

$$r_{t,n} = \sigma_t \frac{1}{\sqrt{N}} s_n z_{t,n}, \tag{1}$$

where $\sigma_t$ is the conditional daily volatility at day $t$, $s_n$ is the intraday volatility periodicity during the period of $r_{t,n}$ and $z_{t,n}$ is an *iid* series of zero mean and unit variance that is independent of the daily volatility $\sigma_t$ and the intraday periodicity $s_n$. The volatility components $\sigma_t$ and $s_n$ are required to be positive. For instance, $\sigma_t > 0$ for all $t$ and $s_n > 0$ for all $n$. The intraday volatility periodicity component $s_n$ is a periodic function at daily frequency. Specifically, $s_{n+jN} = s_n$ for all $j$ and $n$. Thus, the daily return at day $t$ $R_t$, is

$$R_t = \sum_{n=1}^{N} r_{t,n} = \sigma_t \frac{1}{\sqrt{N}} \sum_{n=1}^{N} s_n z_{t,n}. \tag{2}$$

With respect to the model estimation, Andersen and Bollerslev (1997,1998) apply a flexible Fourier transform to estimate the intraday volatility periodicity component $s_n$ and use standard GARCH volatility model to estimate the conditional daily volatility $\sigma_t$. For further details of the Fourier transform, see the appendix of Andersen and Bollerslev (1997).

Under the assumption that the intraday returns are serially uncorrelated, the daily conditional variance is the sum of variances from each time interval. Thus,

$$E\left(\sum_{n=1}^{N} \frac{r_{t,n}^2}{\sigma_t^2}\right) = 1. \tag{3}$$

(2.3) immediately gives that $\sum_{n=1}^{N} s_n^2 / N = 1$. This intuitive specification of intraday volatility has two important features. First, it extends naturally to standard specifications of daily return volatility models with intraday innovations and deterministic daily volatility. Specifically, in the case that $s_n = 1$ for all $n$, the $r_{t,n} = \sigma_t \frac{1}{\sqrt{N}} z_{t,n}$. Consequently, $r_t = \sigma_t \frac{1}{\sqrt{N}} \sum_{n=1}^{N} z_{t,n}$. Thus, the daily return at day $t$ is the sum of product of independent intraday innovations and daily conditional volatility patterns. The innovation component $\frac{1}{\sqrt{N}} \sum_{n=1}^{N} z_{t,n}$ is clearly an *iid* series with zero mean and unit variance.

Besides, the model does not affect the autocorrelation structure of the returns measured at daily frequency. When returns are measured at daily frequency, $R_t$ the daily return at day $t$ is $R_t = \sum_{n=1}^{N} \frac{s_n}{\sqrt{N}} z_t \sigma_t$ where $z_t$ is an *iid* series of zero and unit variance. Thus, the expected absolute return $E|R_t|$ is,

$$E|R_t| = \sum_{n=1}^{N} \frac{s_n}{\sqrt{N}} E|z_t|\sigma_t. \tag{4}$$

Observe that since $E(\sum_{n=1}^{N} s_n^2/N) = 1$, $\sum_{n=1}^{N} \frac{s_n}{\sqrt{N}} \geq 1$ since $\left(\sum_{n=1}^{N} \frac{s_n}{\sqrt{N}}\right)^2 = (\sum_{n=1}^{N} s_n)^2/N \geq \sum_{n=1}^{N} s_n^2/N$. Consequently, the expected daily absolute return is an increasing function of the variation of the intraday periodicity pattern. Let $c = (E|z_t|)^{-2} - 1 > 0$. I immediately have,

$$Corr(|R_t|, |R_\tau|) = \frac{Cov(\sigma_t, \sigma_\tau)}{Var(\sigma_t) + cE(\sigma_t^2)}. \tag{5}$$

(5) clearly suggests that the model does not impose any structural changes to the autocorrelation of the daily returns.

Second, the model gives a qualitative description of the impact of intraday volatility periodicity on the autocorrelation of the absolute intraday returns. Specifically, suppose that $n \geq m$, then

$$Corr(|R_{t,n}|, |R_{\tau,m}|) = \frac{\sum_{j=1}^{N} s_j s_{j-(n-m)} Cov(\sigma_t, \sigma_\tau) + Cov((s_n, s_m) E^2(\sigma_t)}{\frac{Var(\sigma_t)\left(\sum_{j=1}^{N} s_j\right)^2}{N} + cE(\sigma_t^2)\frac{\left(\sum_{j=1}^{N} s_j\right)^2}{N} + E^2(\sigma_t)Var(s)}. \tag{6}$$

Notice that, when $n = m$, equation (2.6) reduces to

$$Corr(|R_{t,n}|, |R_{\tau,n}|) = \frac{Cov(\sigma_t, \sigma_\tau) + \left(\frac{Var(s)}{\left(\sum_{j=1}^{N} s_j\right)^2}\right) E^2(\sigma_t)}{Var(\sigma_t) + cE(\sigma_t^2) + \left(\frac{Var(s)}{\left(\sum_{j=1}^{N} s_j\right)^2}\right) E^2(\sigma_t)}. \tag{7}$$

(6) gives a qualitative description between the intraday periodicity and the daily heteroskedasticity of the absolute returns. The positive covariance of the daily return volatility $Cov(\sigma_t, \sigma_\tau)$ is strong when $t$ and $\tau$ are very close. Consequently, it causes positive dependence in the absolute returns, as the distance between t and $\tau$ grows larger this effect becomes less significant. This implication is consistent with the slow decay of the correlations in Figures 5 and 6.

At the same time, the strong intraday volatility periodicity also has an impact on the correlations of absolute returns. For instance, the autocorrelations of 5-minute return series reaches the lowest value generally at the forty lags, which is exactly half of the trading day. The covariance of the intraday volatility periodicity $Cov((s_n, s_m)$ is expected to reach the minimal at the same time according to the U-shape. Figure 6 suggests that the 10-minute return series exhibit similar characteristics as well. These findings further confirm the correspondence between the qualitative implications of (1) and the autocorrelation structure of the intraday absolute returns.

Although the model of (1) gives a plausible explanation of the autocorrelation structure of the absolute intraday returns and might serve as a starting point for the high frequency volatility modelling, the simplistic intraday volatility specification $\sigma_t \frac{1}{\sqrt{N}} s_n$ implies that the only intraday component in the intraday volatility pattern is $s_n$.

Based on the intraday return volatility model of Andersen and Bollerslev (1997,1998), Engle (2012) proposed the following multiplicative specification for intraday return volatility to provide a more realistic volatility dynamic description,

$$R_{t,n} = \sigma_t \epsilon_{t,n} s_n z_{t,n}, \tag{8}$$

where $\sigma_t$ is the conditional volatility at day $t$, $s_n$ is the intraday volatility periodicity component, $\epsilon_{t,n}$ is the stochastic intraday volatility component with $E(\epsilon_{t,n}^2) = 1$ and $z_{t,n}$ is an *iid* series that follows the normal distribution with zero mean and unit variance. Moreover, the normalized return $y_{t,n} = \frac{r_{t,n}}{\sigma_t s_n}$ is assumed to follow a GARCH (1,1) process. Specifically,

$$y_{t,n}|F_{t,n-1} \sim N(0, \epsilon_{t,n}),$$

$$y_{t,n} = \frac{r_{t,n}}{\sigma_t s_n} = \epsilon_{t,n} z_{t,n},$$

$$\epsilon_{t,n}^2 = \omega + \alpha y_{t,n-1}^2 + \beta \epsilon_{t,n-1}^2.$$

Similarly, given the assumption that the intraday returns are not serially correlated, one immediately has $E\left(\sum_{n=1}^{N} \frac{r_{t,n}^2}{\sigma_t^2}\right) = 1$. Then $E\left(\sum_{n=1}^{N} \frac{r_{t,n}^2}{\sigma_t^2}\right) = \sum_{n=1}^{N} s_n^2 = 1$. Regarding the model estimation, for the intraday volatility periodicity pattern $s_n$, Engle applies a commercially available risk measure to estimate $\sigma_t$ and further the calculates the $s_n$ as the variance of returns in each time interval after deflating by the conditional daily variance $\sigma_t$.

Specifically,

$$E\left(\frac{R_{t,n}^2}{\sigma_t^2}\right) = E\left(\epsilon_{t,n}^2 z_{t,n}^2 s_n\right) = s_n. \tag{9}$$

Note that the model of both (1) and (8) are based on the return series that are sampled from fixed time interval. In order to study the intraday volatility periodicity for the transaction-aggregated return series, I must estimate the intraday volatility periodicity for irregular spaced time interval during the trading day.

### 4.2 Multiplicative component intraday volatility

In this section I present the multiplicative component intraday volatility model. My model is based on the Andersen and Bollerslev (1997, 1998) and Engle (2012). In 1997, Andersen and Bollerslev proposed the intraday volatility model of (1). They expressed the daily conditional variance as the product of intraday and daily volatility. As discussed in section 4.1, this specification successfully describes the periodic structure of the absolute intraday return and can be extended to standard daily volatility model when returns are measured at daily frequencies. This prior specification is proved to be very fruitful. In 1998, Andersen and Bollerslev further add a component to the multiplicative specification of intraday volatility to capture macroeconomic announcements. The model is widely accepted for the modelling of intraday volatility in foreign exchange data. Later in 2012, Engle propose a multiplicative intraday volatility model that express the intraday conditional variance as a product of intraday periodicity, intraday variance and daily variance.

When using tick-by-tick data for an equity with high liquidity like SPY, the component that accounts for macroeconomic announcements in the volatility model is not very practical. E.g., Andersen and Bollerslev (1998) and Rangel (2009). First, most important macroeconomic announcements such as the release of new monetary police happen before the stock market opens. Second, the timing of announcements that are particularly important to equity markets are hard to predict. Third, it is very difficult to capture and measure the market response to the macroeconomic announcement in a circumstance that the duration between two consecutive transactions is less than 0.1s. Finally, asymmetric information and microstructure are generally believed to play a decisive role in the high frequency return dynamics. However, macroeconomic or public announcement dummies cannot account for information arrival through order flow (Engle 2012).

Besides, the intraday volatility models of (1) and (8) are based on returns sampled from fixed time interval. It cannot be applied directly for the transaction-aggregated return series. For instance, consider a tick-by-tick return series $r_{t,n}$, where $r_{t,n}$ represents the $n$th tick-by-tick return at day $t$. In general, $w_{t,n} \neq w_{\tau,n}$ when $t \neq \tau$ where $w_{t,n}$ is the waiting-time between transaction $n$ and transaction $n+1$ at day $t$. Consequently, the periodic component $s_n$ in (1) and (8) should be conditional on both $t$ and $n$. I therefore generalize the model of (8) by Engle (2012) to allow the intraday periodicity component to be conditional on the corresponding waiting-time of $r_{t,n}$. I present our intraday volatility model as the follows.

$$r_{t,n}/\sqrt{w_{t,n}} = \sigma_t \epsilon_{t,n} s_{t,n} z_{t,n}, \tag{10}$$

where $\sigma_t$ is the conditional volatility at day $t$, $w_{t,n}$ is the duration that corresponds to $r_{t,n}$, $s_{t,n}$ is the intraday volatility periodicity component, $\epsilon_{t,n}$ is the stochastic intraday volatility component with $E(\epsilon_{t,n}^2) = 1$ and $z_{t,n}$ is an *iid* series that follows the normal distribution with zero mean and unit variance. $\epsilon_{t,n}$ and $z_{t,n}$ are assumed to be independent of each other. $\sigma_t$, $s_{t,n}$ and $\epsilon_{t,n}$ are strictly positive. That is, $\sigma_t > 0$ for all t, $s_{t,n} > 0$ for all $t, n$, $\epsilon_{t,n} > 0$ for all $t, n$.

It is noticeable that in the specification of (10), I do not consider the duration $w_{t,n}$ as a random variable. It is given exogenously. The incorporation of $w_{t,n}$ in (10) therefore is for the normalization of irregularly spaced returns.

There are several reasons behind this consideration. Frist, joint modelling of durations and volatilities requires an explicit parametric modelling of the arrivals of transactions that is beyond the scope of this paper. Second, the intraday periodic pattern of volatilities and durations are quite different. Adding the intraday periodic component into (10) would further complicate the model specification. Finally, the durations are in general correlated with volatilities. The modelling of durations often involves explanatory variables that are volatility measures. Hence, a multiplicative specification seems to be over simplistic.

Under (10) the daily return $R_t$ at day $t$ is,

$$R_t = \sigma_t \sum_{n=1}^{N_t} \sqrt{w_{t,n}} \epsilon_{t,n} s_{t,n} z_{t,n}, \tag{11}$$

where the $N_t$ represents the total number of tick-by-tick transaction records at day $t$. It is clear that $E(R_t) = 0$. Moreover, the expected value of the squared daily return $E(R_t^2)$ is,

$$E(R_t^2) = \sigma_t^2 \sum_{n=1}^{N_t} E(w_{t,n} \epsilon_{t,n}^2 s_{t,n}^2 z_{t,n}^2) = \sigma_t^2 \sum_{n=1}^{N_t} w_{t,n} s_{t,n}^2. \tag{12}$$

Consequently, $\sum_{n=1}^{N_t} w_{t,n} s_{t,n}^2 = 1$. Without the presence of the intraday periodicity, in which case $s_{t,n} = \frac{1}{\sqrt{N_t}}$ for all $n$, the daily return $R_t$ is

$$R_t = \sigma_t \sum_{n=1}^{N_t} \frac{1}{\sqrt{N_t}} \sqrt{w_{t,n}} \epsilon_{t,n} z_{t,n}. \tag{13}$$

Observe that $\sum_{n=1}^{N_t} \frac{1}{\sqrt{N_t}} \sqrt{w_{t,n}} \epsilon_{t,n} z_{t,n}$ is an independent but not necessarily identical distributed sequence of zero mean and unit variance. I further model the $\epsilon_{t,n}$ by a GARCH (1,1) process, that is

$$y_{t,n} | F_{t,n-1} \sim N(0, \epsilon_{t,n}),$$

$$y_{t,n} = \frac{r_{t,n}}{\sigma_t s_{t,n} \sqrt{w_{t,n}}} = \epsilon_{t,n} z_{t,n},$$

$$\epsilon_{t,n}^2 = \omega + \alpha y_{t,n-1}^2 + \beta \epsilon_{t,n-1}^2 \tag{14}$$

In a summary of the extensive ARCH/GARCH literatures, Bollerslev, Chou, and Kroner (1992) argued that GARCH (1,1) is the most popular model. Moreover, even if the conditional variance specification requires more lag terms, the models are not of a higher order than GARCH (1,2) or GARCH (2,1). In 1990, Nelson gave an explanation of this empirical result. Specifically, efficiency considerations favor models of a lower order since many ARCH/GARCH models could be consistent filters of a particular diffusion process.

With respect to the model estimation, Engle (2012) used a commercially available daily volatility measure that is based on the multifactor risk model of Fabozzi, Jones, and Vardharaj (2002). The use of a daily variance forecasts from the structural analysis allows the intraday volatility model to capture industry factor and common liquidity factors. Nevertheless, the market efficiency theory asserts that such information should be included in the market price. Moreover, given the strong ARCH effect of our daily return of SPY, a GARCH modelling seems to be natural. Thus, I capture the $\sigma_t$ by a GARCH specification. Note that the daily variance component $\sigma_t$ can be estimated by daily realized volatility approaches such as the Engle and Gallo (2006) and Zhang et.al. (2005).

For the intraday periodicity component $s_{t,n}$, Andersen and Bollerslev (1997,1998) applied a flexible Fourier transform to estimate the intraday volatility periodicity component $s_{t,n}$. Their approach has an important advantage that it allows the $s_{t,n}$ to be conditional on *t*. Specifically, the intraday periodic pattern is time varying and is dependent on the conditional daily volatility $\sigma_t$. However, although the periodic pattern of financial returns is a well-known empirical stylized fact, there is no widely accepted economic theory that could stipulate the parametric specification of dynamic structure of the intraday periodicity. As a result, nonparametric procedure seems natural. Thus, I follow the methodology of Engle (2012) and consider $s_{t,n}$ as the variance of returns during its corresponding waiting-time deflated by the daily conditional variance. This specification allows the daily conditional variance to be completely free. Specifically, observe that the specification of (10) implies that

$$E\left(\frac{r_{t,n}^2}{\sigma_t^2}\right) = E\left(w_{t,n}\epsilon_{t,n}^2 s_{t,n}^2 z_{t,n}^2\right) = w_{t,n} s_{t,n}^2. \tag{15}$$

(15) establishes the basis of the estimation of intraday periodicity pattern.

## 5 Intraday volatility components and model estimation

In this section I further explain the reasoning behind of the specification of (10) and discuss the econometrics regarding the estimation of the model. In Section 5.1 I discuss the ARCH effect of the daily returns of SPY and the impact of the ARCH effect on the intraday return series. I also present my modelling for the daily conditional variance $\sigma_t$. In Section 5.2 I present my estimation methodology for the intraday volatility periodicity. In Section 5.3 I discuss the econometrics regarding my estimation.

## 5.1 ARCH effect of daily returns

In this section I discuss the ARCH effect of daily returns and estimate the daily conditional volatility $\sigma_t$. I first present the descriptive statistics of the SPY daily returns from January 3, 2005 to December 31, 2014. There are total 2517 daily observations and overnight effect are not included.

|  | Mean | Variance | Kurtosis | Skewness | Maximum | Minimum | First Lag autocorrelation |
|---|---|---|---|---|---|---|---|
| **Daily returns** | 2.03E-06 | 0.000104 | 14.64993 | -0.52421 | 0.076669 | -0.09421 | -0.08628 |
| **Absolute daily returns** | 0.006581 | 6.02E-05 | 25.97425 | 3.688621 | 0.094207 | 0 | 0.330214 |
| **Squared daily returns** | 0.000103 | 1.46E-07 | 197.7099 | 11.9846 | 0.008875 | 0 | 0.227522 |

**Table 5. Descriptive statistics of daily return series, absolute daily return series and squared daily return series**

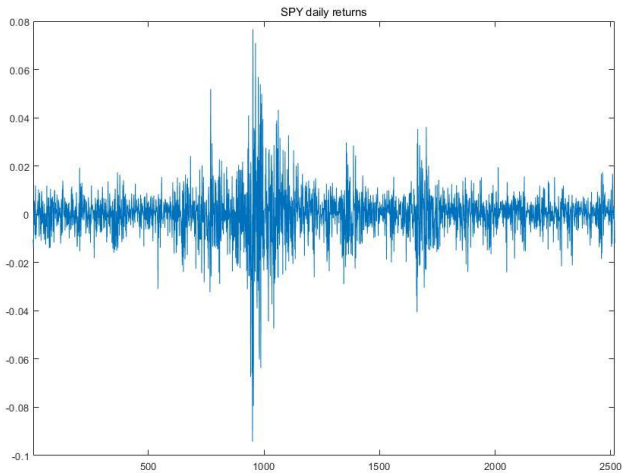

**Figure 7. SPY daily returns from January 3, 2005 to December 31, 2014**

Figure 7 suggests that the daily returns between October 2008 and December 2008 are extremely volatile, which corresponds to the burst of the subprime mortgage financial crisis in 2008. The other busy period following since the end of 2009 corresponds to the European debt crisis. In general, the daily returns fluctuate dramatically around zero. This behavior gives a graphic illustration of the conditional heteroscedasticity.

Besides, although the mean of the daily SPY returns can be safely assumed to be zero, the kurtosis of 14.65 and skewness of -0.52 suggest that the returns are not normally distributed. The daily returns have a small but significant negative first lag autocorrelation. Given the standard significance level of 5%, the lower bound and the upper bound are -0.03986 and 0.03986 respectively. As expected, the daily volatility presents strong volatility clustering phenomenon. The first lag autocorrelations of the absolute daily and squared daily returns are of nonnegligible values and are highly significant. The following correlograms of the absolute daily returns and the squared daily returns provide further evidence.

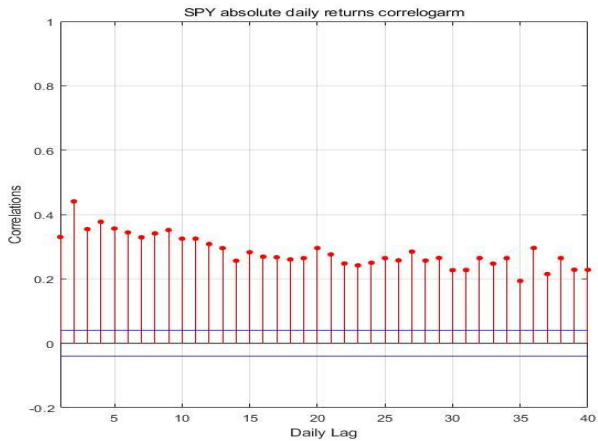

**Figure 8. SPY absolute daily return correlogram**

Figure 8 exhibits the slow decay of the correlation in the volatility of SPY daily returns.

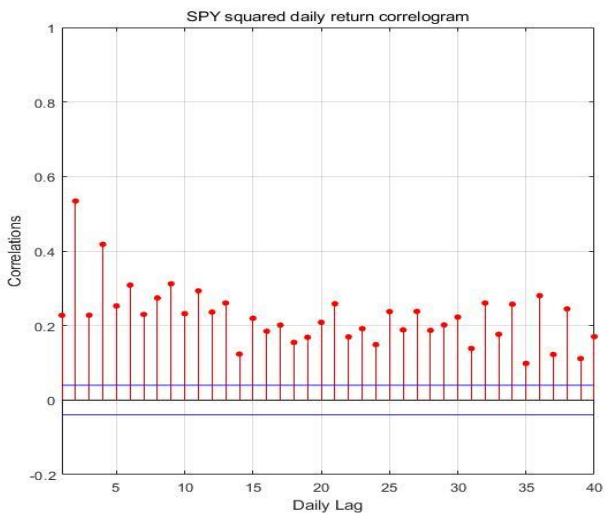

**Figure 9. SPY squared daily return correlogram**

To test the ARCH effect statistically, I apply the ARCH test developed by Engle (1988). Consider an ARCH (p) process,

$$y_t^2 = a_0 + a_i \sum_{i=1}^{p} y_{t-i}^2 + \varepsilon_t.$$

The alternative hypothesis is that there is at least one $a_i \neq 0$ for $i=1,2,3…, q$. The test statistic is the Lagrange multiplier statistic $TR^2$ where $T$ is the sample size and $R^2$ is obtained from fitting the ARCH(p) model via regression. Under the null hypothesis that $a_i = 0$ for all $i$, the asymptotic distribution of the test statistic $TR^2$ is Chi-Square with $p$ degrees of freedom.

A commonly accepted procedure for choosing the number of lags in the ARCH modelling is that one first estimates different lag structures and chooses the one with the minimal Akaike Information Criterion statistics (AIC). However, this procedure in practice usually favors a relatively long lag structure. Thus, I test the null hypothesis that there is no ARCH effect in the daily series against the alternative hypothesis of an ARCH model with two lagged squared innovations that is equivalent to a GARCH (1,1) model locally (Bollerslev 1986). Although the GARCH (1,1) model is not necessarily the preferred model, it still gives a simple and comprehensive approximation to the dependence structure in the autocorrelation of squared returns. The result shows that the test rejects the null hypothesis at the significance level of 1% with a $p$ value of zero for the test statistics. Specifically, the value of the test statistics is 747.5967 and the critical value for the given significance level of 1% is 9.2103.

The pronounced ARCH effect in the previous discussion might not be very surprising since it is widely documented for financial research. However, the question remains that how this ARCH effect observed from daily returns, the aggregation of tick-by-tick returns, influences the intraday volatility process. According to my prior investigation of the ARCH effect on the daily return series, an intraday volatility process with conditional heteroskedasticity seems natural. Nevertheless, the poor performance of ARCH-GARCH models for the modelling of intraday volatility also is widely reported. (Cumby et al. 1993, West and Cho 1995, Figlewski 1995 and Jorin 1995). In order to explicitly explain the intraday volatility process, I first investigate the relation between the daily observed ARCH effect and intraday volatility pattern.

To assess the influence of the ARCH effect observed from daily returns on the intraday returns, we explore the relation between one-step-ahead volatility estimation provided by GARCH models on daily returns and other daily ex post return variability measures calculated from intraday returns. I choose the daily returns from January 3, 2005 to December 31, 2013 to estimate the GARCH model and further compare the one-step-ahead GARCH estimations with volatility measures calculated from intraday returns using sample from January 2, 2014 to December 31, 2014.

For the conditional heteroskedasticity modelling of the daily returns, I choose the MA (1)-GARCH (1,1) model specification. The moving average term MA (1) is included to explain the weak but significant negative first order autocorrelation in Table 5. The model is specified as,

$$R_t = u + \theta \varepsilon_{t-1} + \varepsilon_t,$$
$$\varepsilon_t = \sigma_t z_t,$$
$$\sigma_t^2 = \omega + \alpha \varepsilon_{t-1}^2 + \beta \sigma_{t-1}^2, \quad (16)$$

where $z_t$ is an *iid* sequence and $z_t \sim N$ (0,1). The ARMA is proposed by the thesis of Peter Whittle (1951). The maximum likelihood estimation regarding the ARMA-GARCH process can be found in Ling and Li (1997,1998), McAleer (2003) and Franco (2004). Note that, the GJR-GARCH could be another natural candidate for modelling $\sigma_t^2$ since the daily return series has a skewness of -0.5241. The leverage effect that is the asymmetric contribution of returns with different signs to the variability of the prices is expected to be pronounced when returns are measured at daily frequency.

|   | Values | Standard Error | T statistics | P values |
|---|---|---|---|---|
| $u$ | 0.000272 | 0.000137 | 1.988748 | 0.046729 |
| $\theta$ | -0.06674 | 0.024112 | -2.76797 | 0.005641 |
| $\omega$ | 1.69E-06 | 5.11E-07 | 3.311307 | 0.000929 |
| $\beta$ | 0.873353 | 0.010796 | 80.89544 | 0 |
| $\alpha$ | 0.104926 | 0.009394 | 11.16934 | 5.76E-29 |

**Table 6. Parameter estimation of the MA (1)-GARCH (1,1) model**

Table 6 presents the estimated MA (1)-GARCH (1,1) model. Observe that $\alpha$ is relatively large and $\beta$ is relatively small. The sum of $\beta$ and $\alpha$ is around 0.978 that represents a high level of volatility persistence. The following Figure 2.10 exhibits the plot of standardized residuals and conditional daily variances. Observe the sudden surge of the conditional variance between the interval between 900 to 1000 that corresponds to the dramatic variation of daily returns during the same period in Figure 7.

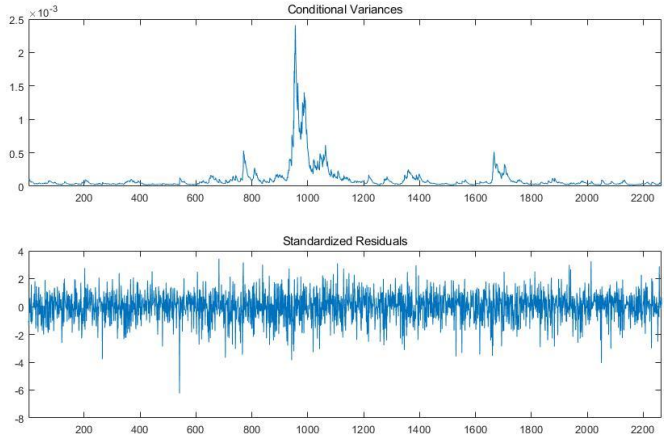

**Figure 10. The conditional variance and standardized residuals of MA (1)-GARCH (1,1) fitting**

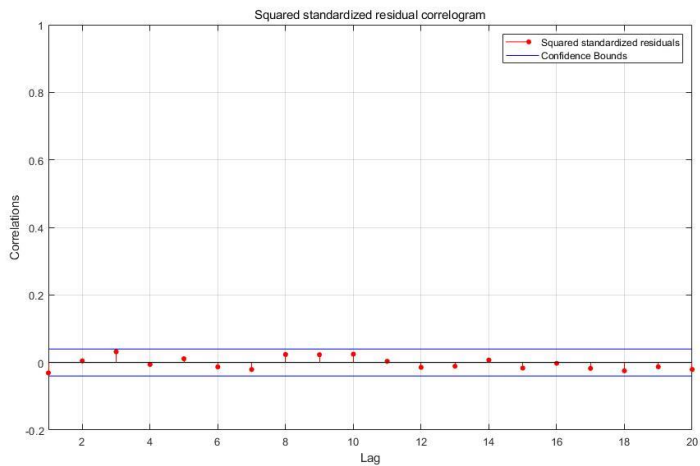

**Figure 11. The correlogram of squared standardized residuals of MA (1)-GARCH (1,1) fitting**

Figure 11 exhibits the autocorrelation function of the squared standardized residuals. The autocorrelations are eliminated as expected. Although the MA (1)-GARCH (1,1) captures the conditional heteroskedasticity successfully as suggested by Figures 10 and 11. Figure 12 suggests the standardized residuals still presents fat tails. Another interesting finding is that, compared with Figure 7, the volatility of SPY daily returns in Figure 10 is much regular. Specifically, the effect of financial crisis in 2008 still stays but other noisy periods start to drop. It might suggest the different effect of intraday events such as mini flash crash and events lasting over days such as the financial crisis in 2008 over the daily volatility pattern of SPY.

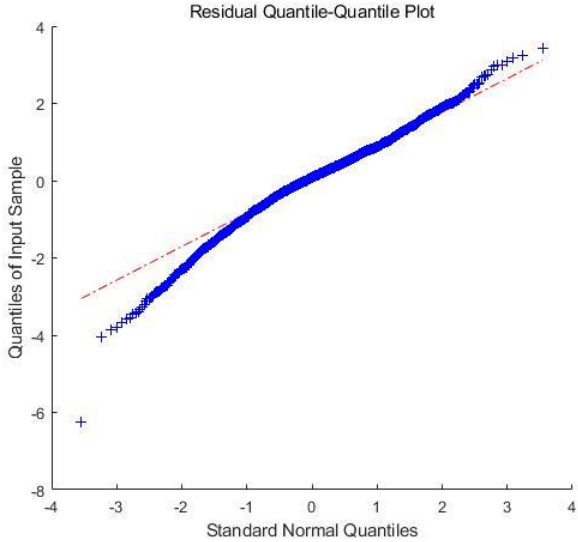

**Figure 10. The Quantile-Quantile Plot of the standardized residuals**

I now turn to the ex post return variability measure that are calculated from intraday returns. The two measures that we choose are the realized variance and the cumulative absolute returns. Specifically, $\sum_{n=1}^{N_t}|r_{t,n}|$ and $\sum_{n=1}^{N_t} r_{t,n}^2$.

Compared with volatility measures that are based on daily returns that cannot capture the variability of intraday prices movement, the realized volatility gives a more realistic estimation since it takes, at the highest frequency, each tick-by-tick observation into account. For instance, the intraday price can fluctuate dramatically over the trading day but ends with a value that is very close to the opening price.

I present the comparison of the one-step-ahead MA (1)-GARCH (1,1) daily forecasts with the absolute daily return $|R_t| = \left[\sum_{n=1}^{N_t} r_{t,n}\right]$, realized variance $\sum_{n=1}^{N_t} r_{t,n}^2$ and cumulative absolute returns $\sum_{n=1}^{N_t}|r_{t,n}|$. Daily returns of SPY from January 3, 2005 to December 31, 2013 are employed to estimate the MA (1)-GARCH (1,1) model. Tick-by-tick transactions from January 2, 2014 to December 31, 2014 are used to calculate the realized volatility and cumulative absolute returns and are aggregated to daily frequency to serve as the out-sample for the GARCH forecasts. All series are normalized to have an average of one.

The following Figure 13 presents the one-step-ahead GARCH (1,1) forecasts and the daily absolute returns $|R_t|$. The daily absolute returns fluctuate arbitrarily arounds the GARCH predictions. It seems intuitively that the GARCH forecasts should have weak explanatory power of the observed daily variability of returns that is measured by, $|R_t| = \left[\sum_{n=1}^{N_t} r_{t,n}\right]$. Further, it might also implicitly suggest that the ARCH effects of the daily returns have negligible impact on the intraday volatility. However, the suggestion is misleading, since

$$|R_t| = \left[\sum_{n=1}^{N_t} r_{t,n}\right] = \left|\sum_{n=1}^{N_t} \ln\left(\frac{P_{t,n+1}}{P_{t,n}}\right)\right| = |lnP_{t,N_t+1} - lnP_{t,1}|. \tag{17}$$

(17) suggests that under the assumption that there is no over night effect $|R_t| = \left[\sum_{n=1}^{N_t} r_{t,n}\right]$ should coincide with the daily absolute returns. In real world, the two concepts are not equivalent for sure. Thus, $|R_t| = \left[\sum_{n=1}^{N_t} r_{t,n}\right]$ only catupres the variablility for aggregated tick-by-tick prices. It actully just takes the first and last tick-by-tick prices into account.

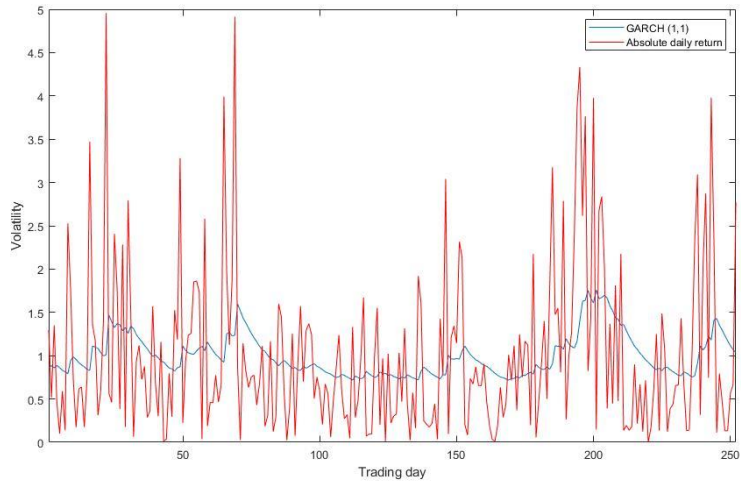

**Figure 13.2 The absolute daily returns and GARCH (1,1) volatility forecasts**

In contrast, Figure 14 suggests the daily GARCH forecasts are much more in line with the ex post daily return variability measure $\sqrt{\sum_{n=1}^{N_t} r_{t,n}^2}$ and $\sum_{n=1}^{N_t} |r_{t,n}|$. These two measures are calculated by tick-by-tick returns and account for the variability of every tick-by-tick return.

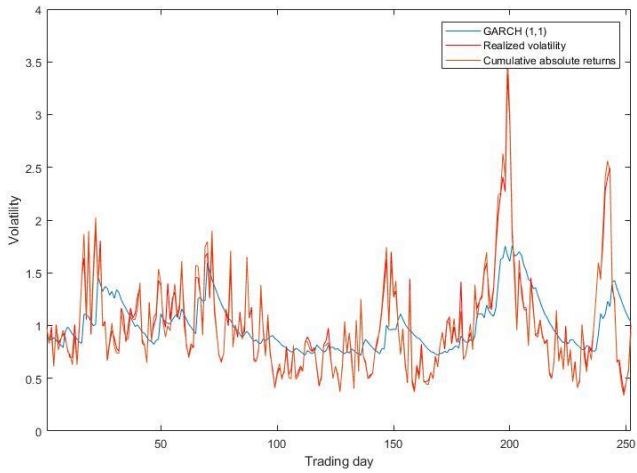

**Figure 14.3 The cumulative absolute returns, realized volatility and GARCH (1,1) forecasts**

To further illustrate this issue, I examine the correlations between the daily GARCH forecasts and the other three series. The results are reported in Table 7.

|  | $\sum_{n=1}^{N_t} |r_{t,n}|$ | P value | $|R_t| = \left|\sum_{n=1}^{N_t} r_{t,n}\right|$ | P value | $\sqrt{\sum_{n=1}^{N_t} r_{t,n}^2}$ | P value |
|---|---|---|---|---|---|---|
| $\sigma_t$ | 0.553814319 | 1.17E-21 | 0.236701 | 1.49E-04 | 0.559359 | 3.80E-22 |
|  | $[\sum_{n=1}^{N_t} |r_{t,n}|]^2$ | P value | $R_t^2$ | P value | $\sum_{n=1}^{N_t} r_{t,n}^2$ | P value |
| $\sigma_t^2$ | 0.54229162 | 1.14E-20 | 0.230797 | 2.19E-04 | 0.548984 | 3.08E-21 |

**Table 7. Correlations between GARCH forecasts and other ex post daily variability measures**

Table 7 confirms the finding from Figures 13 and 14. The correlation between the GARCH forecasts and the realized volatility is around 0.559. That is, 31.25 percent of the variation in the realized volatility can be explained by the MA (1)-GARCH (1,1) model. Besides, the correlation between the GARCH forecast and cumulative absolute returns is also as high as 0.5538. In contrast, the correlation between the GARCH forecast and absolute daily returns is only 0.2367. Thus, only about 6 percent of the variation of the daily absolute returns can be captured by the MA (1)-GARCH (1,1) forecasts.

Moreover, since the GARCH forecasts are based on daily observations, the strong correlation between GARCH forecasts and realized variance cannot be explained by only intraday volatility pattern. No matter what the intraday volatility pattern is, it should be annihilated when returns are measured at daily frequency. Consequently, it suggests that the ARCH effects observed at daily frequency have strong impact on the intraday volatility pattern. As a result, ignoring the pronounced ARCH effect at daily frequency and the corresponding GARCH-ARCH forecasts in intraday volatility modelling inevitably results in the loss of a large percentage of predictable intraday return variability.

According to the analysis and discussion in this section, a misspecification of the intraday volatility process might emerge if we cannot capture the daily ARCH effect. The empirical results of Figures 13, 14 and Table 7 justify the daily conditional component $\sigma_t$ in (10) and support the use of the GARCH prediction to estimate the $\sigma_t$.

### 5.2 Intraday volatility periodicity

In this section, I present our estimation of the intraday volatility periodicity component $s_{t,n}$. In Andersen and Bollerslev (1998), the intraday returns are modelled by the following equation.

$$r_{t,n} = E(r_{t,n}) + \sigma_{t,n} s_{t,n} z_{t,n}, \qquad (18)$$

where the conditional volatility component is $\sigma_{t,n}$, $s_{t,n}$ is the intraday periodicity component and $z_{t,n}$ is an *iid* sequence with zero mean and unit variance. For the estimation of $\sigma_{t,n}$, they use the estimator $\sigma_t/\sqrt{N}$ where $\sigma_t$ is the daily GARCH estimation. Therefore, the intraday periodicity component $s_{t,n}$ is considered as a stochastic process evolving with time and is estimated by flexible Fourier transform. Regarding the functional data analysis of intraday volatility periodicity are in Muller et al. (2007) and Andersen and Bollerslev (1998). Alternatively, Engle (2012) considers the $s_{t,n}$ as constants and grants complete freedom to the daily volatility component $\sigma_t$. Specifically, $s_{t,n} = s_{\tau,n} = s_n$ for all $\tau$ and $t$.

Here, I follow the methodology of Engle (2012) for the estimation of $s_{t,n}$ based on following considerations. First, if I assume that the $E(r_{t,n})$ in (18) to be zero and replace the intraday volatility component $\sigma_{t,n}$ with its estimator $\sigma_t/\sqrt{N}$. The equation now reduces to (1), which is

$$r_{t,n} = 1/\sqrt{N} \sigma_t s_{t,n} z_{t,n}.$$

The intraday component in (18) is now $s_{t,n}$. It evidently should be assumed as a stochastic process. In contrast, recall the specification of intraday returns in (10),

$$r_{t,n}/\sqrt{w_{t,n}} = \sigma_t \epsilon_{t,n} s_{t,n} z_{t,n},$$

where $\epsilon_{t,n}$ is the intraday volatility component and is stochastic. If I consider $s_{t,n}$ as a stochastic process, it is hard to assume that it should be independent of $\epsilon_{t,n}$ and further complicate the picture. In order to explicitly exhibit the intraday return dynamics, it seems natural to assume the periodicity component to be unconditional. Secondly, based on our discussion of realized variance, I should have,

$$E(\sigma_t^2) = E(\sum_{n=1}^{N_t} r_{t,n}^2) = \sum_{n=1}^{N_t} E(w_{t,n}\sigma_t^2 s_{t,n}^2).$$

It intuitively suggests that $s_{t,n}^2$ should be considered as time invariant as well.

I now present the estimation of intraday periodicity component $s_{t,n}$. Since that $E\left(\frac{r_{t,n}^2}{\sigma_t^2}\right) = E(w_{t,n}\epsilon_{t,n}^2 s_{t,n}^2 z_{t,n}^2) = w_{t,n}s_{t,n}^2$, under the assumption that $w_{t,n} = w_{\tau,n}$ for all $t$ and $\tau$ (time-aggregated returns), a natural way to estimate the $s_{t,n}^2$ is

$$\widehat{s_{t,n}^2} = \frac{1}{T}\sum_{t=1}^{T}\frac{r_{t,n}^2}{w_{t,n}\sigma_t^2}. \tag{19}$$

This is reasonable for intraday return series formed from identical time interval (time-aggregated returns). Since my tick-by-tick and transaction-aggregated are not equally spaced in time, in general the $n$th return of day $t$, $r_{t,n}$ and $n$th return of day $\tau$, $r_{\tau,n}$ are not sampled from the same time interval if $t \neq \tau$. I circumvent the difficulty by followings.

Let $x_{t,n}$ represent the time of the $n$th transaction at day $t$. Since $r_{t,n} = \ln(\frac{P_{t,n+1}}{P_{t,n}})$, it can be viewed as the return of day $t$ on period from $x_{t,n}$ to $x_{t,n+1}$. I now consider the intraday volatility pattern during the period $x_{t,n}$ to $x_{t,n+1}$ on each trading day. For another day $k$, let $k(x_{t,n+1}) = \max\{s|x_{k,s} \leq x_{t,n+1}, s = 1,2,3...,\}$, $k(x_{t,n}) = \min\{s|x_{k,s} \geq x_{t,n}, s = 1,2,3...,\}$. Let $S_k(t,n) = \{s|k(x_{t,n}) \leq s \leq k(x_{t,n+1})\}$. The prices at $x_{t,n}$ and $x_{t,n+1}$ on day $k$ are conjectured by taking the average of adjacent two transactions in case that $S_k(t,n) = \emptyset$.

I then define $V_k(t,n) = \frac{1}{S[S_k(t,n)]} \sum_{s \in S_k(t,n)} r_{k,s}^2 / w_{k,s} \sigma_k^2$ where $S[S_k(t,n)]$ is the number of elements in $[S_k(t,n)]$. Further, $\widehat{s_{t,n}^2} = \frac{1}{N} \sum_{k=1}^{N} V_k(t,n)$ is applied to estimate the intraday volatility periodicity pattern during that period. The Matlab code for the estimation is given in Appendix C.

My intraday periodicity estimation procedure allows the shape of the periodic pattern depend on the estimated trading period. It turns out to be important for the tick-by-tick return series and transaction-aggregated return series for which the observations are not equally spaced in time. Besides, my approach uses the full tick-by-tick return series to accurately capture the periodicity.

However, due to the extremely large tick-by-tick sample size, the searching algorithm used to find out $S_k(t,n) = \{s | k(x_{t,n}) \leq s \leq k(x_{t,n+1})\}$ requires very long execution time. Moreover, the random-access memory requirement for the computation is beyond the capacity of my equipment. However, it is possible for traders with high technique capacity to use the tick-by-tick based algorithm to estimate the intraday periodic volatility pattern. Thus, I choose the time-aggregated returns to examine our specification since searching a regular partition of time massively reduce the computational difficulties.

### 5.3 Econometrics regarding estimation

In this section, I discuss statistical properties regarding my model estimation. I estimate the daily conditional volatility component $\sigma_t$ use the MA (1)-GARCH (1,1) model first. Although in principle the parameters of the model in (10) should be estimated simultaneously, I do not adapt this standard procedure according to the following considerations. First, the daily GARCH model estimation generally requires a relatively large sample. However, the corresponding tick-by-tick high frequency data are not available. Second, a tick-by-tick transaction records of several years will have huge number of observations. E.g. My tick-by-tick transaction records of symbol SPY during 01/02/2014-12/31/2014 has 79156264 observations. Consequently, it rises the practical difficulty of storing and computing data of such magnitude in a joint estimation. Finally, since $\sigma_t$ is estimated by MA (1)-GARCH (1,1) without a clear specification in (10), it gives the flexibility to use other daily volatility measures.

Thus, the estimation now is a two-step procedure. The first step I estimate the intraday volatility periodicity component as discussed in Section 5.2. The second step I estimate the GARCH (1,1) model for the stochastic intraday volatility component $\epsilon_{t,n}$. In general, results from a multi-step estimation procedure can be flawed by the errors that are induced by errors in the previous estimation steps. In order to show the estimation are consistent, I apply the generalized methods of moments methodology (GMM).

Let $\psi$ be the $k_1$ parameters estimated at the first step and $\theta$ be the $k_2$ parameters estimated at the second step. Besides, suppose $g_1(\psi)$ and $g_2(\psi,\theta)$ are the $k_1$ and $k_2$ moments conditions which specify the parameters by true moments. Denote the corresponding sample sum as $g_{1,M}$ and $g_{2,M}$. Let $g_M = (g'_{1,M}, g'_{2,M})'$, $\delta = \begin{pmatrix}\psi\\\theta\end{pmatrix}$ and $g(\delta) = \begin{pmatrix}g_1(\psi)\\g_2(\psi,\theta)\end{pmatrix}$. The GMM estimator of the parameter can be presented as

$$\begin{pmatrix}\hat{\psi}\\\hat{\theta}\end{pmatrix} = \arg\min g_M I g_M = \arg\min g_M\, g_M$$

$I$ is the diagonal matrix. In order to solve the above equation, $\psi$ must solve the equation in first row and $\theta$ should solve the equation in the second row given $\hat{\psi}$. According to Newey and McFadden (1994), if $\hat{\psi}$ and $\hat{\theta}$ are consistent estimators of the $\psi$ and $\theta$, under quite general conditions (Newey and McFadden 1994) of $g_M$, the following estimator $\sqrt{M}\begin{pmatrix}\hat{\psi} & \psi\\\hat{\theta} & \theta\end{pmatrix}$ is consistent and asymptotically normal. Specifically, it converges in distribution to the normal distribution $N(0, G^{-1}\Lambda G^{-1'})$ where $G = E(\partial g(\delta)/\partial \delta)$ and $\Lambda = E(g(\delta)g(\delta)')$.

As showed by Hansen (1982) and Engle (2012), the matrix $\sqrt{M}\begin{pmatrix}\hat{\psi} & \psi\\\hat{\theta} & \theta\end{pmatrix}$ can be consistently estimated by using sample averages to replace expectations and using estimation of parameters to replace parameters.

In the context of my estimation, let $r_{t,n}$, $\epsilon_{t,n}$, $\sigma_t$ and $s_{t,n}$ be defined as in equation (10). I have

$$g_{1,M}(\psi) = g_{1,M} = \begin{pmatrix} 1/T \sum_{t,1}^T \frac{r_{t,1}^2}{\sigma_t^2} - s_{t,1}^2 \\ \vdots \\ 1/T \sum_{t,N_t}^T \frac{r_{t,N_t}^2}{\sigma_t^2} - s_{t,N_t}^2 \end{pmatrix},$$

$$g_{2,M}(\hat{\psi}, \theta) = g_{2,M} = 1/T \sum \sum_{t=1}^T \sum_{n=1}^{N_t} \nabla_\theta \left(\log(\epsilon_{t,n}^2) + \left(\frac{r_{t,1}^2}{\sigma_t^2 \widehat{s_{t,n}}^2 \epsilon_{t,n}^2}\right)\right),$$

$$G = \frac{1}{M}\sum \begin{pmatrix} \nabla_\psi g_1 & 0 \\ \nabla_\psi g_2 & \nabla_\theta g_2 \end{pmatrix},$$

$$\frac{1}{M}\sum \begin{pmatrix} g_{1,t}^2 & g_{1,t}g_{2,t} \\ g_{1,t}g_{2,t} & g_{2,t}^2 \end{pmatrix} \to \Lambda. \tag{20}$$

The convergence is in the sense of convergence in probability measure. Moreover, the $\hat{\psi}$ and $\hat{\theta}$ estimated at the first and second stage are consistent estimator for $\psi$ and $\theta$ clearly. Under the assumption of stationarity and ergodicity, Large Number Theorem verifies the consistency of $\hat{\psi}$. Consistency of the GARCH parameters can be found in Lee and Hansen (1994) and Bollerslev (1992).

## 6 Empirical results

In this section I present the empirical results of my intraday volatility specification of (10). I evaluate the specification from mainly two perspectives. First, I analyze the intraday returns normalized by one-day-ahead GARCH forecasts, the intraday returns normalized by one-day-ahead GARCH forecasts and intraday volatility periodicity pattern.

I use the terminology *"normalized returns"* to refer to the intraday returns normalized by one-day-ahead GARCH forecasts. Specifically, $\widetilde{r_{t,n}} = r_{t,n}/\hat{\sigma}_t$. The terminology *"filtered returns"* are used to describe intraday returns normalized by one-day-ahead GARCH forecasts and intraday volatility periodicity pattern. Specifically, $\widehat{r_{t,n}} = r_{t,n}/w_{t,n}\hat{\sigma}_t\widehat{s_{t,n}}$. As explained in previous section 5.2, the de-seasonal technique requires large amount of computation for the transaction-aggregated returns since it is based on searching tick-by-tick transactions records. Thus, I only exhibit the filtered returns for time-aggregated returns. Since those returns are sampled from identical time intervals, the duration $w_{t,n}$ in

(10) can be defined as unit.

Second, I present the model estimation for time-aggregated returns at various frequencies. I compare the estimated parameters with the theoretical aggregational prediction of GARCH (1,1) from Nelson (1992) and analyze the model residuals.

### 6.1 Normalized intraday returns and filtered intraday returns

In this section I present the analysis regarding normalized intraday returns and filtered intraday returns. I divide the analysis into two sections. Section 6.1.1 discusses the analysis of the normalized intraday returns. Section 6.1.2 represents the analysis of the filtered intraday returns.

#### 6.1.1 Normalized intraday returns

Figure 15 presents the one-day-ahead MA (1)-GARCH (1,1) estimates for the period during 01/02/2014-12/31/2014. The details of the estimation are presented in section 5.1. I now present a summary of descriptive statistics for the normalized intraday returns $r_{t,n}/\hat{\sigma}_t$.

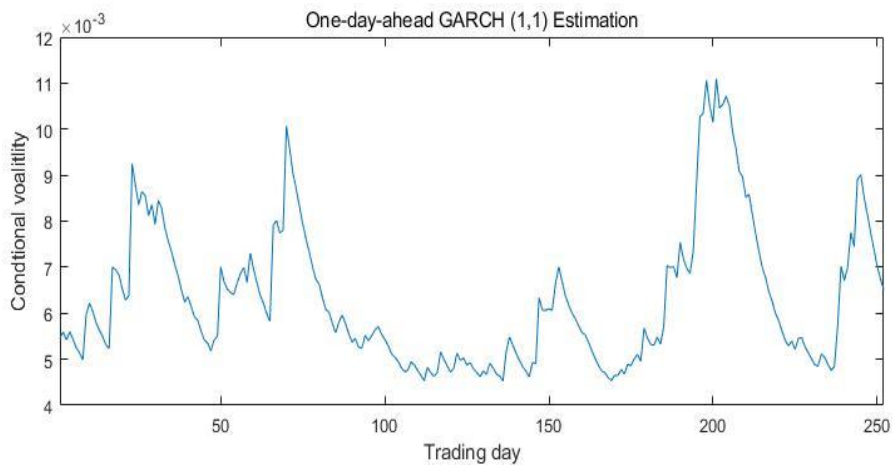

**Figure 15. One-day-ahead MA (1) -GARCH (1,1) estimation**

|  | Sample Size | Mean | Variance | Kurtosis | Skewness | Max | Min | $p_1$ | $(a)p_1$ | $(s)p_1$ |
|---|---|---|---|---|---|---|---|---|---|---|
| 30Second | 196560 | 1.25E-07 | 0.001204 | 167.665 | -0.13508 | 1.385718 | -1.46815 | -0.10463 | 0.312765 | 0.489134 |
| 1Minute | 98280 | 2.51E-07 | 0.002231 | 64.42646 | -0.1912 | 1.408863 | -1.45948 | -0.08058 | 0.282229 | 0.467608 |
| 1.5Minute | 65520 | 3.76E-07 | 0.00321 | 48.6927 | -0.16571 | 1.361122 | -1.49995 | -0.06177 | 0.268761 | 0.453377 |
| 3Minute | 32760 | 7.52E-07 | 0.006195 | 20.9807 | -0.26811 | 1.245272 | -1.63514 | -0.04921 | 0.240136 | 0.378386 |
| 5Minute | 19656 | 1.25E-06 | 0.0095 | 6.321807 | -0.14592 | 0.731673 | -0.83181 | -0.01585 | 0.19892 | 0.15415 |
| 10Minute | 9828 | 2.51E-06 | 0.018818 | 5.993268 | -0.22319 | 0.814763 | -0.98106 | -0.01946 | 0.190533 | 0.125991 |
| 15Minute | 6552 | 3.76E-06 | 0.027404 | 5.769621 | -0.23229 | 0.945254 | -1.09734 | 0.003031 | 0.181626 | 0.105811 |
| 30Minute | 3276 | 7.52E-06 | 0.055366 | 5.137176 | -0.25934 | 1.091958 | -1.15388 | 0.008075 | 0.152081 | 0.107696 |
| 78Minute | 1260 | 1.96E-05 | 0.146711 | 4.904028 | -0.39886 | 1.448271 | -1.88065 | 0.068372 | 0.167124 | 0.112384 |

Table 8. The normalized time-aggregated returns

|  | Sample Size | Mean | Variance | Kurtosis | Skewness | Max | Min | $p_1$ | $(a)p_1$ | $(s)p_1$ |
|---|---|---|---|---|---|---|---|---|---|---|
| T=400 | 198261 | -1.9E-06 | 0.001155 | 1308.24 | 0.189044 | 3.307417 | -3.29112 | -0.0962854 | 0.253815 | 0.499309 |
| T=800 | 98942 | -6.1E-06 | 0.002064 | 210.4466 | 0.008546 | 2.210617 | -2.19975 | -0.047105 | 0.182147 | 0.495312 |
| T=1200 | 65893 | -9.5E-06 | 0.003171 | 372.8296 | 0.021205 | 3.31828 | -3.31285 | -0.0569775 | 0.194002 | 0.497701 |
| T=2400 | 32876 | 0.000106 | 0.005926 | 113.042 | 2.467613 | 3.35631 | -0.55958 | 0.00530582 | 0.064171 | 0.001178 |
| T=4000 | 19663 | 1.99E-05 | 0.009473 | 3.529551 | -0.04754 | 0.557197 | -0.58379 | 0.00794604 | 0.07018 | 0.088125 |
| T=8000 | 9763 | -2.2E-05 | 0.019155 | 3.4 | -0.05032 | 0.755934 | -0.53138 | -0.0011895 | 0.052102 | 0.050723 |
| T=12000 | 6468 | 0.000195 | 0.028742 | 3.314235 | -0.15596 | 0.696386 | -0.64206 | -0.0049805 | 0.059878 | 0.055918 |
| T=24000 | 3162 | 0.000689 | 0.057648 | 3.233583 | -0.15545 | 0.891024 | -0.91855 | -0.0108197 | 0.032436 | 0.032172 |
| T=62000 | 1165 | -0.00216 | 0.161005 | 3.056574 | -0.13479 | 1.343841 | -1.22803 | 0.01098296 | 0.082668 | 0.060503 |

Table 9. The normalized transaction-aggregated returns

Tables 8 and 9 give the descriptive statistics of the time-aggregated returns and transaction-aggregated at various frequencies, where $p_1$, $(a)p_1$ and $(s)p_1$ are the first lag autocorrelations of the normalized returns, absolute normalized returns and squared normalized returns respectively.

Although I again detect little interests in the sample mean, the skewness and kurtosis have intriguing features. Regarding the kurtosis, it turns out that when returns are measured at the highest frequencies, the normalized time-aggregated returns have much smaller kurtosis compared with the transaction-aggregated returns. For instance, the kurtosis of the normalized 30-second, 1-minute, 1.5-minute and 3-minute returns series are 167.67, 64.43, 48.69 and 20.9807 respectively. Meanwhile, the kurtosis of the corresponding normalized T-400, T-800, T-1200 and T-2400 transaction aggregated returns series are 1308.24, 210.45, 372.83 and 113.04 respectively. In other words, transaction-aggregated returns are more fat-tailed compared with time-aggregated returns at the highest frequencies. However, when returns are measured at relatively low level of intraday frequency, the normalized transaction-aggregated returns exhibit much regular kurtosis compared with corresponding normalized time-aggregated returns. For example, the kurtosis of the corresponding normalized T-4000, T-8000, T-12000 and T-24000 transaction aggregated returns series are 3.53, 3.4, 3.33 and 3.23 respectively. However, the corresponding normalized 5-minute, 10-minute, 15-minute and 30-minute return series have kurtosis of 6.32, 5.99, 5.77 and 5.13 respectively. Observe the sharp drop of kurtosis between the T-2400 and T-4000 normalized transaction aggregated returns.

Regarding the skewness, note that the normalized T-4000 and T-8000 return series have skewness that are very close to 0. In contrast, the skewness of normalized time-aggregated returns varies and present no obvious dependence on the return frequencies.

Another important empirical finding comes from the last three columns in Tables 8 and 9. First, both normalized intraday time-aggregated and transaction-aggregated returns are approximately uncorrelated, although there is a small, but significant, negative first order autocorrelations at the highest frequencies. This is explained the bid-ask bounce as mentioned previously.

Second, in general the normalized intraday returns still exhibit significant volatility clustering feature. For instance, the normalized 1-minute, 1.5-minute and 3-minute squared returns have the first lag autocorrelations of 0.4676, 0.4533 and 0.3784 respectively. Meanwhile, the first lag autocorrelations of these three normalized time-aggregated returns are -0.0806, -0.0618 and -0.0492 that are neglectable from the economic perspective. The normalized transaction-aggregated returns exhibit similar characteristics. For example, the normalized T-800 and T-1200 squared returns have very strong positive first lag autocorrelations of 0.4953 and 0.4978 and meanwhile the first lag autocorrelation of the normalized T-800 and T-1200 return series are –0.0481 and -0.0570. This finding suggests the intraday returns exhibit strong ARCH effect even after normalizing by ARCH-GARCH daily volatility forecasts. This implicitly supports our specification of (10). However, this intraday ARCH effect generally dies out when return frequencies decrease as suggested by the monotonically decreasing values of the last two columns in Tables 8 and 9.

Third, compared with the normalized time-aggregated return series for which the intraday ARCH effect is significant even at very low intraday frequency, the ARCH effect of normalized transaction-aggregated return series vanishes very quickly. For instance, when the number used to aggregate the tick-by-tick transactions exceeds 2400, the first lag autocorrelation of normalized squared transaction-aggregated returns and normalized absolute transaction-aggregated are of very small values that can be assumed zero plausibly.

Under (10), The normalized time-aggregated intraday returns $\frac{r_{t,n}}{\sigma_t} = \epsilon_{t,n} s_{t,n} z_{t,n}$. Thus, in addition to the discussed conditional heteroskedasticity, it should present the intraday periodicity pattern as well. I here present the correlogram of the normalized 5-minute absolute return series and the normalized 10-minute absolute return series for up to five trading days.

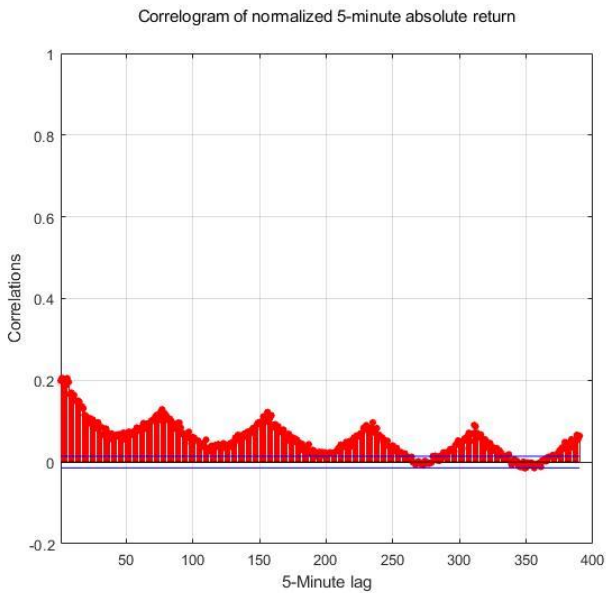

**Figure 4.** The correlogram of normalized 5-minute absolute returns

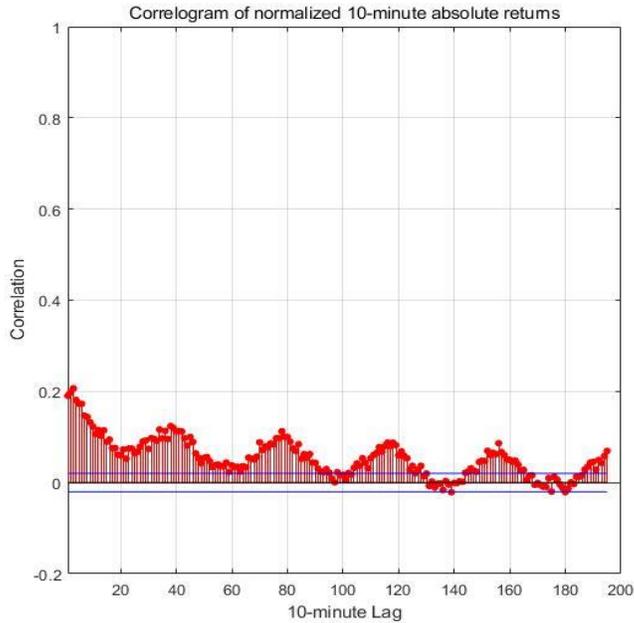

**Figure 17. The correlogram of normalized 10-minute absolute returns**

Figures 16 and 17 exhibit strong periodic pattern as expected. Compared with the Figures 5 and 6 in section 3, a salient feature is that the normalized intraday returns now no longer constantly possess significant autocorrelations for the lags up to five trading days.

### 6.1.2 The filtered intraday returns

In this section, I discuss the filtered intraday returns $\widehat{r_{t,n}} = r_{t,n}/\hat{\sigma}_t\widehat{s_{t,n}}$. I first present a graphic illustration of our estimation of $s_{t,n}$. To be consistent with the previous description, I use the 5-minute time intervals to present the intraday periodicity.

Figure 18 presents the expected U-shape of intraday periodicity. It starts at the highest level when market opens and then gradually decreases with time. It reaches the lowest value at the forty 5-minute trading interval that is around 12:50. After that, it increases with time until the market is closed.

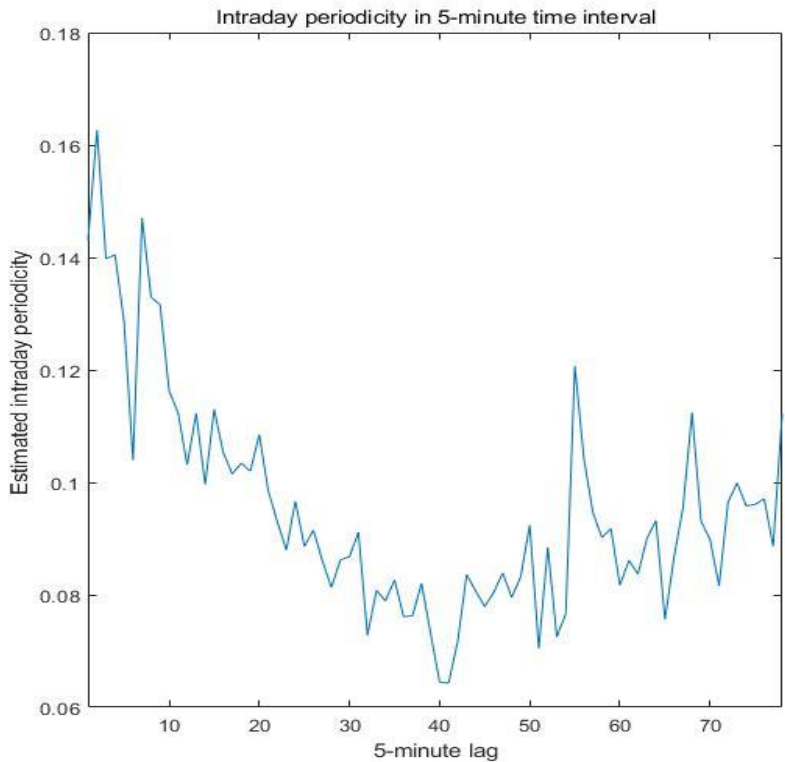

**Figure 18. The estimated intraday periodicity for each 5-minute time interval**

|  | Sample Size | Mean | Variance | Kurtosis | Skewness | Max | Min | $p_1$ | $(a)p_1$ | $(s)p_1$ |
|---|---|---|---|---|---|---|---|---|---|---|
| **30Second** | 196560 | -0.00037 | 1.000005 | 16.04269 | -0.04466 | 15.14489 | -15.1776 | -0.04421 | 0.237787 | 0.39142 |
| **1Minute** | 98280 | -0.00047 | 1.00001 | 12.22964 | -0.08725 | 14.24929 | -14.3882 | -0.05103 | 0.228836 | 0.346531 |
| **1.5Minute** | 65520 | -0.00073 | 1.000015 | 12.90129 | -0.08684 | 13.81001 | -13.7784 | -0.0321 | 0.217655 | 0.354665 |
| **3Minute** | 32760 | -0.00078 | 1.00003 | 10.25224 | -0.09803 | 12.36469 | -12.5788 | -0.03602 | 0.205242 | 0.321281 |
| **5Minute** | 19656 | -0.00078 | 1.00005 | 5.63634 | -0.13308 | 6.100498 | -8.11604 | -0.02155 | 0.171075 | 0.145198 |
| **10Minute** | 9828 | -0.00118 | 1.0001 | 5.424702 | -0.18769 | 5.90282 | -6.74803 | -0.02201 | 0.169528 | 0.125574 |
| **15Minute** | 6552 | -0.00212 | 1.000148 | 5.408844 | -0.24594 | 5.388117 | -7.40249 | 0.009641 | 0.160309 | 0.099098 |
| **30Minute** | 3276 | -0.00411 | 1.000288 | 4.959931 | -0.27933 | 4.463494 | -5.19666 | 0.012625 | 0.143373 | 0.108887 |
| **78Minute** | 1260 | -0.00165 | 1.000792 | 4.869385 | -0.34463 | 4.633105 | -4.41664 | 0.069233 | 0.181823 | 0.128609 |

**Table 10.1 The filtered time-aggregated returns**

Table 10 summaries the descriptive statistics regarding the filtered time-aggregated returns at different frequencies. Observe the variance now is almost unity for filtered time-aggregated returns as specified under (10). The filtered time-aggregated returns still present excessive high kurtosis and negative skewness, suggesting that the unconditional distribution is not normal. An interesting finding is that the filtered time-aggregated returns now present little evidence for autocorrelations. Specifically, the first lag autocorrelation for the 30-second, 1-minute, 1.5-minute, 3-minute and 5-minute return series are now -0.044, -0.051, -0.032, -0.036 and -0.022 respectively. These values are still significant but of small magnitude. In contrast, the squared filtered time-aggregated return series still present strong autocorrelations as suggested by the last column of Table 10. This implies the existence of conditional heteroskedasticity.

I carry out the ARCH test (Engle 1988) to verify the conditional heteroskedasticity as in Section 5.1. Specifically, we test the null hypothesis that there is no ARCH effect in the filtered time-aggregated return series against the alternative hypothesis of an ARCH model with two lagged squared innovations that is equivalent to a GARCH (1,1) model locally.

|  | P values | Value of $TR^2$ | Critical value of $TR^2$ |
| --- | --- | --- | --- |
| 30Second | 0 | 32328.71024 | 9.210340372 |
| 1Minute | 0 | 12087.05103 | 9.210340372 |
| 1.5Minute | 0 | 8450.709077 | 9.210340372 |
| 3Minute | 0 | 3402.702996 | 9.210340372 |
| 5Minute | 0 | 728.5254549 | 9.210340372 |
| 10Minute | 0 | 303.1955374 | 9.210340372 |
| 15Minute | 0 | 123.6137445 | 9.210340372 |
| 30Minute | 0 | 113.2577976 | 9.210340372 |
| 78Minute | 0 | 86.10606377 | 9.210340372 |

**Table 11. ARCH test of filtered time-aggregated returns**

As the suggested by Table 11, the null hypothesis of no ARCH effect is rejected at all frequencies. The significance level is 0.01. In other words, the time-aggregated returns after normalization by daily conditional variance and intraday periodicity still present strong conditional heteroskedasticity. Finally, I present the correlogram of the absolute filtered 5-minute and 10-minute time-aggregated return series for up to five days.

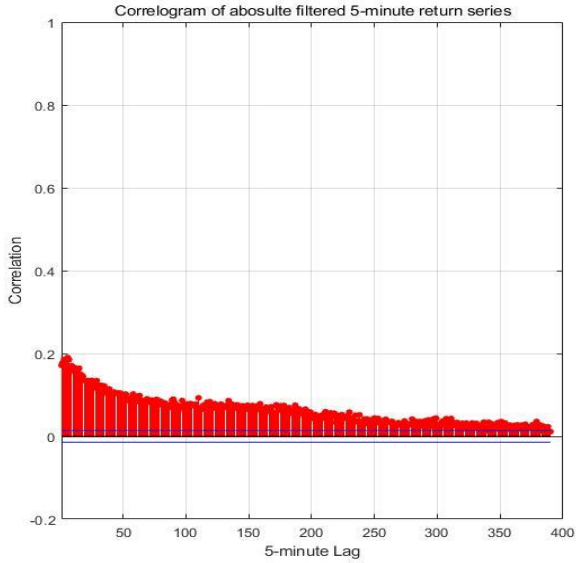

**Figure 19. The correlogram of filtered 5-minute absolute returns**

The autocorrelations of filtered 5-minute absolute returns decreases now very slowly through the lags since the intraday periodic pattern has been excluded by $\widehat{r_{t,n}} = r_{t,n}/\widehat{\sigma_t}\widehat{s_{t,n}}$. This is in line with autocorrelation structure of GARCH model. The U-shape in the Figures 15 and 16 now no longer exists. It suggests that my estimation of the intraday volatility periodicity is effective. The following Figure 20 is equally telling.

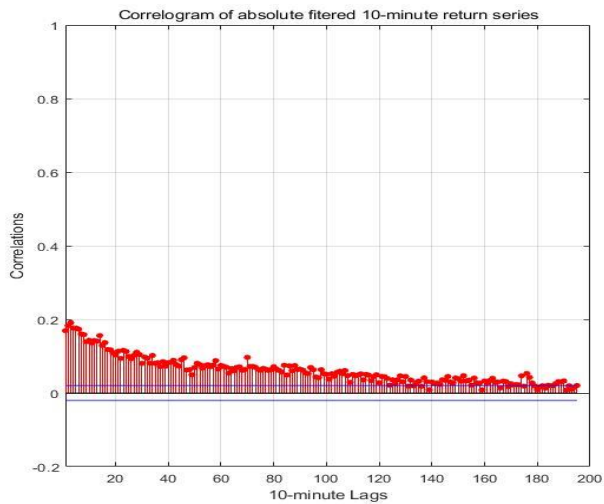

**Figure 5. The correlogram of filtered 10-minute absolute returns**

## 6.2 Results of GARCH estimation and volatility persistence

In this section I present the results of the GARCH estimation. I first exhibit the inadequacy of the standard GARCH modelling for describing intraday volatility process. I then present our GARCH estimation of the filtered time-aggregated return volatility. I focus on the comparison between the estimates and the theoretical aggregation results from Nelson (1990,1992). Finally, I present the model residuals using the example of 5-minute and 10-minute filtered returns.

There are vast literatures regarding the modelling of high frequency return dynamics using GARCH (1,1) model. Thus, I first examine the effectiveness of the GARCH (1,1) in modelling raw intraday time-aggregated returns. In order to capture the significant first order autocorrelation, I choose the MA (1)-GARCH (1,1) specification.

$$R_{t,n} = u + \theta \varepsilon_{t,n-1} + \varepsilon_{t,n},$$

$$\varepsilon_{t.n} = \sigma_{t.n} z_{t.n},$$

$$\sigma_{t.n}^2 = \omega + \alpha \varepsilon_{t,n-1}^2 + \beta \sigma_{t,n-1}^2, \qquad (21)$$

where $E_{t,n-1}[(\varepsilon_{t,n}^2)] = \sigma_{t.n}^2$ is the conditional variance and $z_{t.n} \sim N(0,1)$ and is *iid*.

The estimation is based on the quasi-maximum likelihood methods. Given the intraday conditional heteroskedasticity discussed previous, it seems natural to choose the GARCH (1,1) specification for the intraday variance modelling since it usually represents a reasonable approximation. Besides, estimating the GARCH (1,1) model across all time-aggregated return frequencies yields meaningful comparisons between estimated parameters. In order to describe the volatility persistence implied by the GARCH (1,1) parameters, I use the "half-life" that is the number of periods the process takes for half of the expected reversion backs to the unconditional variance. Specifically, $-ln2/ln(\alpha + \beta)$. The "mean lag" is also supplied as an additional measure of the volatility persistence. i.e. $\alpha(1 - \alpha - \beta + \alpha\beta)^{-1}$.

Nelson (1990,1992) and Drost and Nijman (1993) provide a framework to assess the GARCH parameter estimates at various sampling frequencies. Specifically, consider the GARCH (1,1) at the daily frequency. Let $\alpha^{(1)}$ and $\beta^{(1)}$ be the parameters for the GARCH (1,1) daily return model, then the $\alpha^{(t)}$ and $\beta^{(t)}$ that are parameters for the GARCH model for the $t$-day returns should suffice the equation,

$$\alpha^{(t)} + \beta^{(t)} = (\alpha^{(1)} + \beta^{(1)})^t. \tag{22}$$

Equation (22) immediately implies that the estimated $t$-day GARCH half-life should be related with daily GARCH half-life as the following equation,

$$-ln2/\ln(\beta^{(t)} + \alpha^{(t)}) = -ln2\frac{1}{\ln(\alpha^{(1)}+\beta^{(1)})^t} = -\frac{1}{t}\frac{ln2}{ln(\alpha^{(1)}+\beta^{(1)})}. \tag{23}$$

As a result, the estimated half-life of GARCH models of return series with different sampling frequencies should be stable if they are normalized by corresponding frequencies. Besides, as showed by Nelson (1990, 1992), the relations hold under general conditions of GARCH (1,1), no matter whether the model is mis-specified at some frequencies or not. Empirical evidence for the relations between daily and lower frequency GARCH estimation can be found in Drost and Nijman (1993), Drost and Werker (1996) and Bollerslev (1997).

| | $\alpha$ | Standard error | $\beta$ | Standard error | $\alpha + \beta$ | Half-life | Median lag |
|---|---|---|---|---|---|---|---|
| 30-Second | 0.157 | 0.0065542 | 0.918 | 0.0022418 | 1.075 | ∞ | ∞ |
| 1-Minute | 0.139047 | 0.0015039 | 0.681 | 0.0051612 | 0.820047 | 3.493795158 | 0.506280349 |
| 1.5-Minute | 0.114605 | 0.0011804 | 0.434 | 0.0053909 | 0.548605 | 1.73177985 | 0.343036541 |
| 3-Minute | 0.290064 | 0.0067264 | 0.195893 | 0.0029211 | 0.485957 | 2.881572988 | 1.524343333 |
| 5-Minute | 0.335584 | 0.0106232 | 0.250198 | 0.0068206 | 0.585782 | 6.480334013 | 3.368093719 |
| 10-Minute | 0.318509 | 0.0143567 | 0.472546 | 0.0092005 | 0.791055 | 29.57271835 | 8.860894976 |
| 15-Minute | 0.278234 | 0.0100601 | 0.575654 | 0.014528 | 0.853889 | 65.82402449 | 13.62654377 |
| 30-Minute | 0.160019 | 0.0110557 | 0.7568 | 0.0083571 | 0.916819 | 239.4433497 | 23.49961427 |
| 78-Minute | 0.148918 | 0.0177689 | 0.791006 | 0.0153911 | 0.939924 | 872.6371124 | 65.30338637 |

**Table 10. GARCH estimation of intraday time-aggregated returns**

The half-life and median-lag in Table 12 are converted into unit minute. Consider the daily MA (1)-GARCH (1,1) estimation that gives the $\alpha = 0.104 \ and \ \beta = 0.873$, the results of Drost and Nijman (1993) now suggest that the intraday returns should follow weak GARCH (1, 1) processes with $\alpha + \beta$ approaching 1 and $\alpha$ approaching 0 as the frequencies increase. However, the $\alpha$ and $\beta$ in Table 12 behave erratically since the sum of the two parameters does not exhibit such tendency clearly. Moreover, the half-life and median-lag present dramatic variations along with the sample frequencies. It is very clear that a direct use of GARCH (1,1) specification of intraday return volatility is seriously in doubt. I now present the results of GARCH modelling of the filtered time-aggregated return $R_{t,n}/\widehat{\sigma_t s_{t,n}}$ in further details.

| | $\omega$ | Stand error | T statistics | $\alpha$ | Standard error | T statistics | $\beta$ | Standard error | T statistics | $\alpha + \beta$ | Half-life | Median lag |
|---|---|---|---|---|---|---|---|---|---|---|---|---|
| 30second | 0.00471 | 6.04E-05 | 77.95481167 | 0.04074 | 0.000236967 | 171.939 | 0.95588 | 0.000240364 | 3976.801 | 0.9966 | 102.422 | 0.4813303 |
| 1minute | 0.00575 | 0.000113 | 50.86504525 | 0.04562 | 0.000533077 | 85.58773 | 0.94968 | 0.00054553 | 1740.836 | 0.9953 | 147.25 | 0.9500191 |
| 1.5minute | 0.00592 | 0.0001692 | 34.97214726 | 0.05346 | 0.000738496 | 70.29338 | 0.94265 | 0.000760529 | 1276.446 | 0.9961 | 266.826 | 1.4772541 |
| 3minute | 0.0087 | 0.0003331 | 26.11107358 | 0.05715 | 0.001174732 | 48.64809 | 0.93564 | 0.001247306 | 750.1245 | 0.9928 | 287.116 | 2.8251007 |
| 5minute | 0.00936 | 0.0005619 | 16.65479684 | 0.06235 | 0.002083684 | 29.92373 | 0.9295 | 0.002230022 | 416.8098 | 0.9918 | 423.339 | 4.7158355 |
| 10minute | 0.01609 | 0.0013179 | 12.20932696 | 0.08228 | 0.003900577 | 21.09331 | 0.90385 | 0.00415363 | 217.6043 | 0.9861 | 496.054 | 9.3240088 |
| 15minute | 0.02301 | 0.002541 | 9.05468895 | 0.111 | 0.006779362 | 16.37279 | 0.87125 | 0.007439406 | 117.1129 | 0.9822 | 580.461 | 14.546339 |
| 30minute | 0.03857 | 0.0043299 | 11.19373674 | 0.10985 | 0.009813608 | 11.19374 | 0.85459 | 0.01109267 | 77.04099 | 0.9644 | 574.33 | 25.46061 |
| 78minute | 0.11523 | 0.0195757 | 5.886293674 | 0.18905 | 0.028076156 | 6.733532 | 0.70614 | 0.033930637 | 20.81114 | 0.8952 | 488.296 | 61.877751 |

**Table 13. GARCH estimation of filtered intraday time-aggregated returns**

Table 13 suggests that the estimation now is much more in line with theoretical predictions. First, the half-life estimated from various frequencies are relatively stable now. For instance, the half-life estimated from the filtered 5-minute, 10-minute, 30-minute and 78-minute return series are 423.34, 496.05, 580.46, 574.33 and 488.30 respectively. Second, the parameters $\alpha$ and $\beta$ are strikingly regular as the theoretical prediction. Observe now $\alpha$ monotonically decreases and meanwhile $\beta$ monotonically increases as the sample frequencies increase. Since the daily MA (1)-GARCH (1,1) estimation that gives the $\alpha = 0.104 \ and \ \beta = 0.873$, according to (22), I should observe that $\alpha + \beta$ approaches one and $\alpha$ approaches zero as the frequencies increases, which is perfectly matched by the empirical results.

I now present the residual correlogram of the filtered 5-minute return series and filtered 10-minute return series.

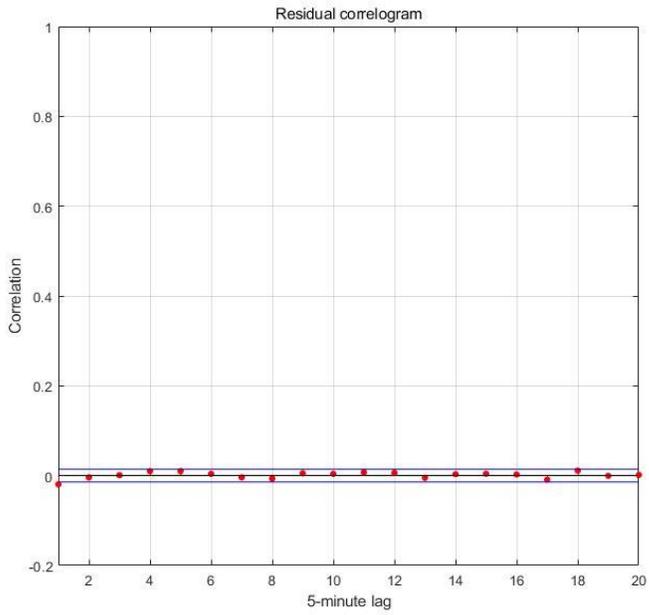

**Figure 01. Residuals correlogram of the filtered 5-minute returns**

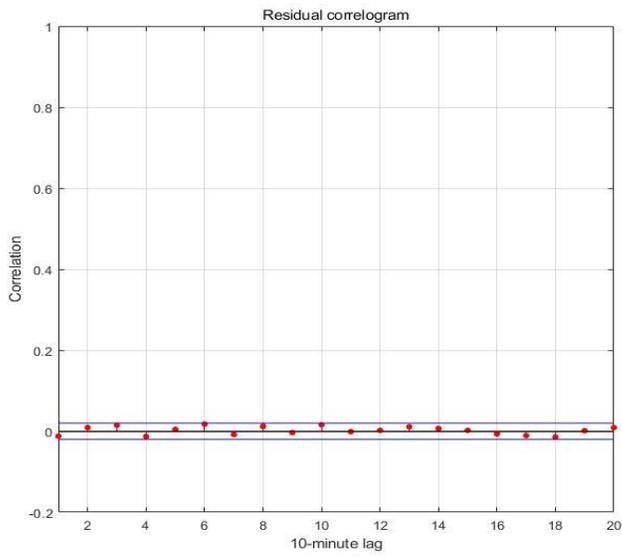

**Figure 02. Residuals correlogram of the filtered 10-minute returns**

Figures 21 and 22 suggest that the intraday conditional heteroskedasticity in Figures 19, 20 and Table 11 is successfully captured by the GARCH (1,1) model. Figure 23 gives the residual Q-Q plot of the filtered 5-minute returns, the residuals still present the fat-tailed property that is widely identified by high frequency financial research.

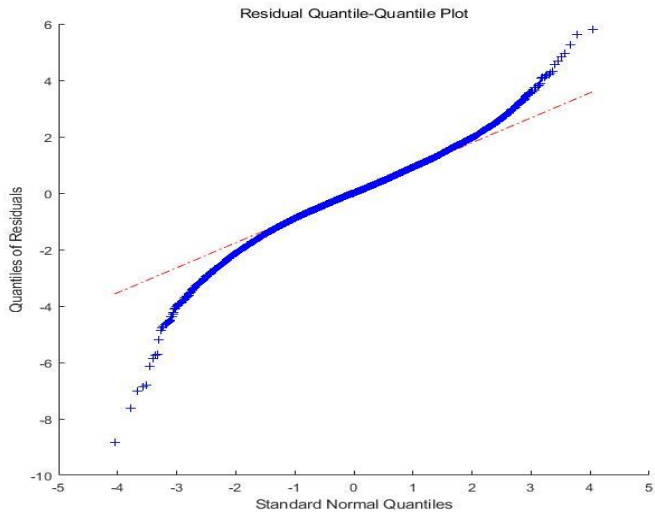

**Figure 03. The Q-Q plot of filtered 5-minute residuals**

## 7   Conclusion

In this paper I analyze the intraday volatility pattern of the SPY on NYSE. I show that standard time series modelling techniques might draw erroneous conclusions due to the distortion induced by strong intraday volatility periodicity. In order to draw meaningful results from the high frequency intraday data, the modelling of the intraday periodicity pattern is necessary. Based on the tick-by-tick transaction records, I develop a non-parametric intraday periodicity modelling methodology that can be used in the data set where observations are not equally spaced in time. However, the estimation procedure for intraday volatility periodicity requires large computation capacities and durations are given exogenously. Further research could be carried out to jointly model the durations and intraday volatility. I also examine the relation between the daily and intraday conditional heteroskedasticity. I find out the intraday conditional heteroskedasticity cannot be fully explained by the long-memory characteristics of daily observations. I prove that the combination of intraday volatility periodicity and intraday conditional heteroskedasticity helps to explain the intraday volatility pattern.

In the analysis of the intraday return dynamics, I find out that a key element that relates closely to the estimation of periodicity and the modelling of instantaneous volatility is the duration between consecutive transactions.

# Appendix A.

The classic paper of Stephens in 1974 presents a detailed discussion with respect to the empirical distribution function statistics such as the statistics $D$ (derived from $D^+$ and $D^-$), $W^2$, V, $U^2$ and $A^2$. Given some random sample $x_1, x_2, x_3, \ldots, x_n$ and the null hypothesis $H_0$ that the random sample comes from a distribution with distribution function $F(x)$, the EDF statistics compare the $F(x)$ with the empirical distribution function $F_n(x)$. In our case, we estimate the $u$ and $\sigma^2$ by the average of the sample $\hat{x}$ and $\hat{\sigma}^2 = \Sigma_i(x_i - \hat{x})/(n-1)$.

The sample are supposed to be indexed by values in ascending order, $x_1 \leq x_2 \leq x_3 \leq \ldots \leq x_n$. The test procedures are organized as follows.

(a) Estimate parameters $u$ and $\sigma^2$ as discussed previously.

(b) Calculate $z_i$'s as $z_i = F(x_i)$, where $F(x)$ is the distribution function of normal distribution with $u$ and $\sigma^2$ estimated in (a).

(c) Calculate $D$, $W^2$, V, $U^2$ and $A^2$ as follows:

1. The Kolmogorov statistics $D$, $D+$ and $D-$:

$$D^+ = max_{1 \leq i \leq n} \left[ \left(\frac{i}{n}\right) - z_i \right];$$

$$D^- = max_{1 \leq i \leq n} \left[ -\left(\frac{i-1}{n}\right) + z_i \right];$$

$$D = \max(D^+, D^-).$$

2. The Cramér–von Mises statistics $W^2$:

$$W^2 = \Sigma_{i=1}^{n}[z_i - \frac{(2i-1)}{2n}]^2 + 1/12n.$$

3. The Kuiper statistic V:

V= $D^+ + D^-$.

4. The Waston statistics $U^2$:

$$U^2 = W^2 - n\left(\sum_{i=1}^{n} \frac{z_i}{n} - \frac{1}{2}\right)^2.$$

5. The Andersen-Darling statistics $A^2$:

$$A^2 = -\frac{\{\sum_{i=1}^{n}(2i-1)[lnz_i+\ln(1-z_{n+1-i})]\}}{n} - n.$$

I produce my table of critical values by Monte-Carol simulation. The null hypothesis $H_0$ that the sample comes from a normal distribution is rejected if the values of the statistics are larger than the critical values given corresponding significance levels. The test is the traditional upper-tail test.

| Statistics | Significance level | | |
|---|---|---|---|
| | 5% | 2.5% | 1% |
| $D$ | 0.895 | 0.955 | 1.035 |
| $V$ | 1.489 | 1.585 | 1.693 |
| $W^2$ | 0.126 | 0.148 | 0.178 |
| $U^2$ | 0.116 | 0.136 | 0.163 |
| $A^2$ | 0.787 | 0.918 | 1.092 |

**Table 1.10(a) Critical values of EDF statistics for the null hypothesis of normal distribution.**

# Appendix B.

The standard Central Limit Theory is the cornerstone of probability theory and statistics. It asserts that:

*If $\{X_t\}$ be a sequence of identically distributed independent random variables with $E(X_t) = u$ and $Var(X_t) = \sigma^2$ for all t. Then, $\lim_{t \to \infty} P\left(\frac{\sum_{t=1}^{n} X_t - nu}{\sqrt{n}\sigma} \leq x\right) = \Phi(x)$, where $\Phi$ is the cumulative distribution function of the standard normal distribution.*

In many economic circumstances, the *i.i.d* assumption cannot be assumed to hold easily. Thus, an issue of both practical and theoretical importance is that:

Assume that $\{X_t\}_{t \in Z}$ is a process with zero mean $E(X_t) = 0$ and finite second moment $E(X_t^2) < \infty$. Define the partial sum $S_n = \sum_{t=1}^{n} X_t$ and its normalized variance $s_n^2 = \frac{Var(S_n)}{n}$. Whether or not we can apply the Central Limit Theorem to have $\lim_{n \to \infty} P(S_n \leq x\sqrt{ns_n^2}) = \Phi(x)$ where $\Phi$ is the cumulative distribution function of the standard normal distribution. Moreover, if the central limit theorem holds, what is the possible convergence rate. Most of the literatures regarding the convergence rate use the Kolmogorov metric as the metric of underlying $\{X_t\}$. Define $\Delta_n(x) = |P(S_n \leq x\sqrt{ns_n^2}) - \Phi(x)|$ and further $\Delta_n = \sup\{\Delta_n(x) | x \in R\}$. Hörmann (2009) provides the bounds for normal approximation error $\Delta_n$ for dependent $\{X_t\}$. An important class of dependent $\{X_t\}$ is the ARCH/GARCH process.

In the last decades, the ARCH/GARCH model are widely applied for the modelling of time-varying volatility. In 1997, Duan introduces a general functional form of GARCH model, the so-called augmented GARCH (1,1) process, that contains many existing GARCH models as special cases. The augmented GARCH (1,1) process is given by:

$$y_t = \sigma_t \varepsilon_t,$$
$$\Lambda(\sigma_t^2) = c(\varepsilon_{t-1})\Lambda(\sigma_{t-1}^2) + g(\varepsilon_{t-1}),$$

where $\Lambda$, $c$ and $g$ are real-valued measurable functions and $\{y_t\}_{t \in Z}$ is a random variable and $\{\varepsilon_t\}_{t \in Z}$ is an *i.i.d* sequence. In order to solve $\sigma_t^2$, the definition requires the existence of $\Lambda^{-1}$. Aue et al. (2006) discuss the necessary and sufficient conditions for the augmented GARCH (1,1) model to have a strictly stationary and non-negative solution for $\sigma_t^2$. For the special case of GARCH (1,1),

$$\sigma_t^2 = \omega + \beta \sigma_{t-1}^2 + \alpha y_{t-1}^2 = \omega + [\beta + \alpha \varepsilon_{t-1}^2] \sigma_{t-1}^2 .$$

It requires that $E(log|\beta + \alpha \varepsilon_0^2|) < 0$. Hörmann (2008) analyzes the dependence structure and asymptotical properties of the augmented GARCH process and shows that *m*-dependent approximations $\{Y_{tm}\}$ to the original sequence of $\{Y_t\}$ can be obtained. Specifically, $||Y_{tm} - Y_t||_2 < const \cdot \rho^m$ ($\rho < 1$), where $||\cdot||_2$ is the $L^2$ norm. He then uses the *m*-dependent approximations to construct the proof of convergence to normal distribution and corresponding convergence rate. Details of the proof can be found in Hörmann (2008).

Based on this result, consider the ARMA (*p*,*q*)-GARCH (1,1) model:
$$x_t = \phi_1 x_{t-1} + \cdots + \phi_p x_{t-p} + \theta_1 y_{t-1} + \cdots + \theta_q y_{t-q},$$
$$y_t = \sigma_t \varepsilon_t,$$
$$\sigma_t^2 = \omega + \beta \sigma_{t-1}^2 + \alpha y_{t-1}^2.$$

If a strictly stationary and casual solution of the equation $x_t = \phi_1 x_{t-1} + \cdots + \phi_p x_{t-p} + \theta_1 y_{t-1} + \cdots + \theta_q y_{t-q}$ exists, Brockwell and Davis (1991) show that the solution can be presented as a linear process $\sum_i \psi_i y_{t-i}$ where the coefficients $\psi_i$ exponentially decay as the lags. Then the convergence of the distribution of normalized sums of $x_t's$ to the normal distribution can be verified. According to Hörmann (2009), if $p \in (2,3]$ moments exist for $x_t$, $\Delta_n = O((logn)^{p-1} n^{1-\frac{p}{2}})$.

However, our tick-by-tick return data presents many erratic statistical features that cannot be explained by the ARMA-GARCH models. The fit of ARMA-GARCH models using tick-by-tick returns gives poor results. The required stationarity is violated since the fit gives the value of $\alpha + \beta$ that exceeds 1. More importantly, the raw tick-by-tick returns are seriously affected by the market microstructural issues such as "split-transactions" which we cannot analyze since the bid and ask prices are not included in the dataset. Thus, a theoretical modelling of the tick-by-tick price dynamics seem to be impractical.

# Appendix C

The following Matlab program "tickstnadardize" estimates the periodicity pattern for given trading interval. In practice, it is used with other programs jointly. For instance, if I want to filter the transaction-aggregated returns T-400, I will use the T-400 return series and corresponding tick-by-tick table in a for-loop to find out its periodicity pattern. The for-loop will search the whole tick-table. For time-aggregated return series, the computation is much simpler since the for-loop will be executed for fixed time interval.

```
function [r1] =tickstandardize(Ticktable,tradetimes,r)

%  tradetimes and r are one day's data for a specific day, Ticktable is

%  252*4 cell table for tick transactions

n=length(r); % number of returns in r series

for i=1:n

s2=tradetimes(i+1);

s1=tradetimes(i); % trade interval is [s1 s2] for return r(i)

v=[];

for j=1:252

 tickprices=Ticktable{1}{j};

 ticktimes=Ticktable{2}{j};

 transactions=[]; % This variable is for the transactions that occurs between s1 and s2 on day j

  for k=1:length(ticktimes)

    if ticktimes(k)>=s1 && ticktimes(k)<=s2

    transactions=[transactions;tickprices(k)];
```

```matlab
        elseif ticktimes(k)>s2
            break
        end
    end
    standardizer=log(transactions(2:end))-log(transactions(1:end-1));
    v(j)=sum(standardizer.^2);   % use realized variance to estimate the periodicity
end
sv=(sum(v)/252)^(1/2);
r1(i)=r(i)/sv;
end
end
```

The following Matlab program "tfiteration" uses the conditional daily variance and intraday periodicity to normalize the intraday transaction-aggregated returns.

```
function [p,t,r,adjustr,ar] = tfiteration(data,tickt,tickp,T,v)

% This function uses the conditional daily variance and intraday

% periodicity to normalize the returns.

% data is the spyfiltered data,tickt,tickp are the coresppongding

% ticktable, T is the frequency index

dayindex=datasort(data{2});

for i=1:length(dayindex)-1

dayprices=data{4}(dayindex(i)+1:dayindex(i+1));

daytradetimes=data{3}(dayindex(i)+1:dayindex(i+1)); % extract the day data

[p{i},t{i},r{i},~]=tsprices(dayprices,daytradetimes,T);   % get the intraday transaction-aggregated returns.

s=[];

for j=1:length(r{i})

t1=t{i}(j);

t2=t{i}(j+1); % [t1,t2] is the timeinterval of jth intraday return

prices1=pricelock(t1,tickp,tickt);

prices2=pricelock(t2,tickp,tickt);

ticksquare=log(prices2./prices1).^2;

s(j)=mean(ticksquare./v); %s(j) is actually s^2(j)

end
```

```
adjustr{i}=(r{i}/sqrt(v(i)))./sqrt(s);
end
ar=[];
for i=1:length(dayindex)-1
ar=[ar, adjustr{i}];
end
end
```

This Matlab program "pricelock" searches the tick-by-tick transaction records for transactions occurring in specific time interval.

```matlab
function [prices] = pricelock(timing,tickp,tickt)

for i=1:length(tickp)

t=tickt{i};    %get the ith day tick time table
n=length(t);   %n is the length of transactions

right=t(t>=timing);
left=t(t<=timing);

if isempty(right)
   pright=tickp{i}(end);
else
   pright=tickp{i}(n-length(right)+1);
end

if isempty(left)
   pleft=tickp{i}(1);
else
   pleft=tickp{i}(length(left));
end

prices(i)=(pright+pleft)/2;
end
```